\documentclass{emulateapj}

\newcommand{\ldl}{$\lambda/{\Delta}{\lambda}$}
\newcommand{\teff}{T$_{eff}$}
\newcommand{\ki}{\ion{K}{1}}
\newcommand{\csi}{\ion{Cs}{1}}
\newcommand{\nai}{\ion{Na}{1}}
\newcommand{\rbi}{\ion{Rb}{1}}

\newcommand{\cai}{\ion{Ca}{1}}
\newcommand{\caii}{\ion{Ca}{2}}

\newcommand{\lii}{\ion{Li}{1}}
\newcommand{\tii}{\ion{Ti}{1}}

\newcommand{\meth}{CH$_4$}
\newcommand{\wat}{H$_2$O}

\newcommand{\kms}{km~s$^{-1}$}
\newcommand{\cmss}{cm~s$^{-2}$}

\slugcomment{Submitted to ApJ 20 July 2006; Accepted for publication 3 October 2006}

\shorttitle{2MASS Ultracool Subdwarfs}
\shortauthors{Burgasser, Cruz \& Kirkpatrick}

\begin{document}

\title{Optical Spectroscopy of 2MASS Color-Selected Ultracool Subdwarfs}

\author{Adam J.\ Burgasser}
\affil{Massachusetts Institute of Technology, Kavli Institute for Astrophysics and Space Research,
Building 37, Room 664B, 77 Massachusetts Avenue, Cambridge, MA 02139; ajb@mit.edu}
\author{Kelle L.\ Cruz\altaffilmark{1}}
\affil{Department of Astrophysics, 
American Museum of Natural History, Central Park West at 79th Street, New York, NY 10024; kelle@amnh.org}
\and
\author{J.\ Davy Kirkpatrick}
\affil{Infrared
Processing and Analysis Center, M/S 100-22, California Institute
of Technology, Pasadena, CA 91125; davy@ipac.caltech.edu}

\altaffiltext{1}{NSF Astronomy and Astrophysics Postdoctoral Fellow.}

\begin{abstract}
We present Gemini GMOS and Magellan LDSS-3 
optical spectroscopy for seven ultracool subdwarf
candidates color-selected from the Two Micron All Sky Survey.
Five are identified as late-type subdwarfs,
including the previously reported sdM9.5 SSSPM 1013-1356 
and L subdwarf 2MASS 1626+3925, and a new sdM8.5 2MASS 0142+0523.
2MASS 1640+1231 exhibits spectral features intermediate
between a late-type M dwarf and subdwarf, similar to the previously
identified  high proper motion star SSSPM 1444-2019, and we classify
both sources as mild subdwarfs, d/sdM9.
2MASS 1227-0447 is a new ultracool extreme subdwarf,
spectral type esdM7.5. 
Spectral model fits yield metallicities that are consistent with 
these metallicity classifications.  Effective temperatures 
track with numerical
subtype within a metallicity class, although they are not equivalent 
across metallicity classes.
As a first attempt 
to delineate subtypes in the L subdwarf regime we classify 2MASS 1626+3925
and the previously identified 2MASS 0532+8246 as sdL4 and sdL7, 
respectively, to reflect their similarity
to equivalently classified, solar metallicity 
L-type field dwarfs over the 7300--9000 {\AA} region.
We also detail preliminary criteria for distinguishing L subdwarf
optical spectra as a roadmap for defining this new spectral class.
The strong TiO bands and \ion{Ca}{1} and \ion{Ti}{1} lines
in the spectrum of 2MASS 1626+3925
provide further evidence that condensate formation
may be inhibited in metal-deficient L subdwarfs. 
We conclude with a compendium of currently known, optically
classified ultracool subdwarfs.
\end{abstract}


\section{Introduction}

Ultracool subdwarfs are metal-deficient, very low mass stars and brown
dwarfs with late spectral types \citep{mecs13}.  
They are the metal-poor analogs of ultracool
dwarfs, spectral types M7 and later \citep{kir97}, and represent the low temperature 
({\teff} $\lesssim$ 3000~K; Leggett et al. 2000)
extensions of the M subdwarf (sdM; [M/H] $\sim$ -1.2) and extreme subdwarf classes
(esdM; [M/H] $\sim$ -2.0; Gizis 1997; Gizis \& Reid 1997; 
see also Mould \& McElroy 1978 and Hartwick, Cowley \& Mould 1984).
Like ultracool dwarfs, ultracool subdwarfs
exhibit complex spectral energy distributions dominated by
strong, overlapping molecular absorption bands; numerous 
neutral metal line features; and red spectral continua
(e.g.~\citet{mou76,bes82,lie87,giz97,leg98,leg00,cus06}).  
Their metal deficiency is 
reflected by enhanced metal hydride and weakened
metal oxide absorption bands at optical wavelengths
\citep{mou76,giz97}, and near-infrared colors that
are unusually blue due to enhanced
collision induced H$_2$ absorption \citep{lin69,sau94,bor97}.
Cool and ultracool subdwarfs typically exhibit 
halo kinematics (${\langle}V{\rangle} = -202$ km s$^{-1}$; Gizis 1997), 
and were presumably formed early in the
Galaxy's history.  With their extremely long nuclear burning lifetimes,
these low mass objects are
important tracers of Galactic structure and chemical enrichment history 
and are representatives of the first generations of star formation.

Ultracool subdwarfs also encompass the new spectral class of L subdwarfs, 
metal-poor analogs to L-type field dwarfs (e.g., Kirkpatrick et al.\ 1999).
These objects are of particular interest for studies of low temperature,
low mass stars and brown dwarfs, and they span masses down to, and possibly below, the
Hydrogen burning minimum mass \citep{me0532}.
L dwarfs and subdwarfs appear to share several common
spectral traits, including strong, pressure-broadened 
\ion{Na}{1} and \ion{K}{1} alkali resonance
lines at optical wavelengths,
prominent FeH and CrH bands, and strong {\wat} 
absorption at near-infrared wavelengths.
L subdwarfs betray their metal deficiency by their 
enhanced metal hydride bands, 
blue near-infrared colors ($J-K$ $\approx$ 0, compared to 1.5--2.0 for 
most field L dwarfs; \citet{kir00}) and
the absence of CO absorption at 2.3 $\micron$, differences
analogous to those observed
between M dwarfs and subdwarfs.  However, the spectrum
of the coolest L subdwarf identified to date, 2MASS J05325346+8246465 
\citep[hereafter 2MASS 0532+8246]{me0532}\footnote{Source designations 
in this article are abbreviated in the manner 2MASS~hhmm$\pm$ddmm;
the suffix conforms to IAU nomenclature convention and is the sexigesimal Right
Ascension (hours and minutes) and declination (degrees and arcminutes)
at J2000 equinox. Full designations are provided
in Table~\ref{tab_targets}.},
exhibits relatively strong TiO absorption, unexpected
in a very cool, metal-poor dwarf; while a second L subdwarf, LSR 1610-0040
\citep{lep1610} exhibits a blend of L dwarf-like and M subdwarf-like
features in its spectrum \citep{cus06,rei06}.
These unexpected properties indicate that much remains to be learned
about these objects.

Over the past few years, the number of known ultracool subdwarfs has surged,
due largely to new proper motion surveys based on photographic $R$-
and $I$-band imaging (e.g., SUPERBLINK, \citet{lep02,lep05};
SuperCOSMOS Sky Survey [SSS], \citet{ham01a,ham01b,ham01c}).
These surveys exploit the high space
velocities of halo subdwarfs 
and their relative brightness at red wavelengths.
In addition, several late-type subdwarfs, including
the two latest-type L subdwarfs now known, 
have been serendipitously identified with 
the Two Micron All Sky Survey \citep[hereafter 2MASS]{skr06}
based on their unusual photometric colors and offset 
optical counterparts (due to proper motion).
Because very late-type subdwarfs such as 2MASS 0532+8246 are
often invisible in red photographic plate images, 
they may be more readily 
identified through wide-field near-infrared imaging, by photometric
color selection and/or proper motion.

In an effort to find
new ultracool subdwarfs, we have obtained
optical spectra for seven candidate sources
color-selected from the 2MASS database.
These observations confirm five as ultracool subdwarfs,
including a new sdM8.5 2MASS 0142+0523; a late-type
mild subdwarf similar to the high proper motion star SSSPM 1444-2019 \citep{sch1444}, 
and a new ultracool extreme subdwarf, the esdM7.5 2MASS 1227-0447.
In $\S$~2 we describe the selection of our candidates
and our optical spectroscopic observations.
In $\S$~3 we present and describe the optical spectra, deriving
atomic line strengths and radial velocities.
Classifications for all seven sources are presented in $\S$~4,
as well as distance estimates and kinematics.
Spectral model fits to the optical data are described in $\S$~5,
and resulting {\teff} and [M/H] determinations analyzed.
In $\S$~6 we discuss our results in the context of 
cool and ultracool subdwarf classification and condensate formation in L
dwarf atmospheres.
Results are summarized in $\S$~7.

\section{Observations}

\subsection{Source Selection}

\begin{figure}
\centering
\epsscale{0.8}
\includegraphics[width=0.8\textwidth]{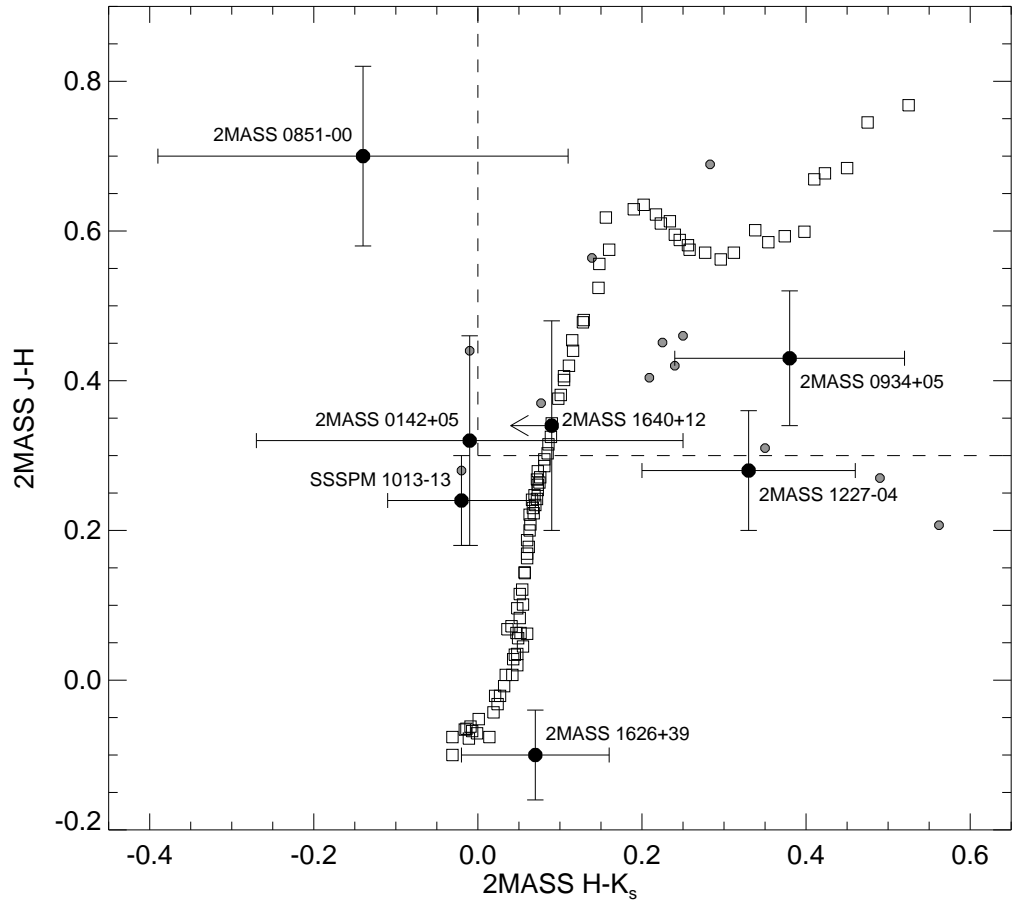}
\caption{2MASS near-infrared color-color diagram for the sources
in our sample (large filled circles with error bars).  Also shown are
mean colors for field B2-M9.5 dwarfs (open squares; R.\ Cutri 2005,
private communication) and previously identified ultracool
subdwarfs and extreme subdwarfs (small filled circles).
The color selection limits for the original 2MASS 
T dwarf search sample
are indicated by dashed lines.
\label{fig_color}}
\end{figure}

The seven objects observed in this study and their empirical properties
are listed in Table~\ref{tab_targets}.  All 
were initially identified in the 2MASS Working Database point source catalog
during the course of a color-selected search for T dwarfs \citep{mewide1},
brown dwarfs that exhibit
{\meth} absorption bands in their near-infrared spectra
\citep{me02a,geb02}.
The color selection criteria for this sample 
target sources with blue near-infrared colors
($J-H$ $\leq$ 0.3 or $H-K_s$ $\leq$ 0) 
and no optical counterpart within 5$\arcsec$ in the USNO-A2.0 catalog 
\citep[effectively $R-J \gtrsim 5$]{mon98}.
These criteria have
also been shown to select late-type, high proper motion subdwarfs,
as strong H$_2$ absorption leads to blue $J-K_s$ colors while 
the motion of the source can result in significant angular displacement 
between the 2MASS and photographic plate images \citep{mewide3}.
Figure~\ref{fig_color} displays a near-infrared color-color diagram
of our candidates and previously identified ultracool subdwarfs,
illustrating how these sources generally lie below the M dwarf 
track\footnote{Note that revised photometry from the 2MASS All Sky Catalog \citep{skr06}
has moved two of our targets, 2MASS 0934+0523 and 2MASS 1640+1231, 
outside of the original near-infrared color selection criteria.}
\citep[R.\ Cutri, 2005, priv.\ comm.]{bes88}.  Clearly, the T dwarf
color selection criteria are not ideally suited for identifying
ultracool subdwarfs, as several known sources lie within the
excluded color space.  Nevertheless, 
several proper motion subdwarfs have been 
recovered in the 2MASS sample, including the recently reported
late-type M subdwarfs LSR 0822+1700 \citep[esdM6.5]{lep0822} and
SSSPM 1013-1356 \citep[esdM9.5]{sch1013}, and the L subdwarfs
2MASS 0532+8246 \citep{me0532} and 2MASS 1626+3925 \citep{me1626}.  
SSSPM 1013-1356 and 2MASS 1626+3925 are re-examined 
in this study.

Figure~\ref{fig_finders} displays $R$-band (from the 
SERC and POSS-II photographic sky surveys; \citet{har81,can84,rei91,mor92}) 
and $J$- and $K_s$-band images (from 2MASS) for 
2MASS 0142+0523, 2MASS 0851-0005, 2MASS 0934+0536, 2MASS 1227-0447 and 2MASS 1640+1231.
Equivalent finder charts for SSSPM 1013-1356
and 2MASS 1626+3925 are given in \citet{sch1013} and \citet{me1626}, respectively.
For three of these sources (excluding 2MASS 0851-0005 and 
2MASS 0934+0536), faint counterparts
can be seen at slightly offset positions in the $R$-band plates, taken 6-14 years
prior to the 2MASS observations.  
These counterparts were verified by the absence of a near-infrared source
at the optical position and consistency in optical/near-infrared colors
as compared to other known late-type subdwarfs (cf.~$\S$~6.3).
An $I_N$-band counterpart to 2MASS 0851-0005
is also seen in UKST photographic infrared
plates (not shown).\footnote{An $I_N$-band counterpart is also seen in POSS-II infrared
plates of 2MASS 0934+0536, but were taken within the same month of the 2MASS observations,
obviating the opportunity to determine proper motion for this source.  There may be a faint $R$-band counterpart in Figure~\ref{fig_finders}, but it is not listed in the SSS database.}  
For these sources, we measured proper motions by
comparing positional data from the SSS and 2MASS catalogs, assuming 
typical astrometric uncertainties of 0$\farcs$3 for each catalog \citep{ham01a,cut03}.  
For 2MASS 1227-0447,
photographic images spanning 44.7 yr were available with POSS-I $R$-band
plates \citep{abe59}.  The corresponding motions and position angles are listed
in Table~\ref{tab_targets}.  With the exception of 2MASS 0851-0005, all of the
measured proper motions are relatively large 
($\mu \gtrsim 0{\farcs}5$~yr$^{-1}$).

\subsection{GMOS Spectroscopy}

Five sources in our sample, 2MASS 0142+0523, 2MASS 0851-0005, 2MASS 0934+0536,
SSSPM 1013-1356 and 2MASS 1626+3925, were observed with the 
Gemini Multi-Object Spectrograph \citep[hereafter GMOS]{hoo04}, mounted
on the 8m Gemini North Telescope,
during queue scheduled time between August and November 2004.
A log of observations is provided in Table~\ref{tab_log}.
GMOS is a grating-based spectrograph utilizing an  
array of three 2048$\times$4608 EEV CCD detectors with 13.5 $\micron$ pixels.
For our observations, we employed the 0$\farcs$75-wide longslit and R400 grating 
with central wavelengths of 7970 and 8030 {\AA},
yielding 5900-10100 {\AA} spectra with a resolution {\ldl} $\sim$ 1300.  Dispersion
on the chip was 0.67 {\AA}~pixel$^{-1}$.  The OG515 filter was used
to suppress stray light from shorter wavelengths.
Each source was observed in two slit positions for each of the tuned central
wavelengths, yielding four separate exposures with individual integration times
of 450 (SSSPM 1013-1356) or 900~s each.
Standard baseline calibration data were also acquired, 
including flat field and CuAr
lamp exposures (interspersed between the science observations),
and observations of the DAw flux standard G~191-B2B \citep{mas88,mas90}
obtained on 8 October 2004 (UT) using the same instrumental setup
(slit, grating and central wavelength offsets).  

All data were downloaded
from the Gemini Science Archive\footnote{GSA; see
\url{http://www2.cadc-ccda.hia-iha.nrc-cnrc.gc.ca/gsa/index.html}.} and 
reduced with the IRAF\footnote{IRAF is distributed by the National Optical
Astronomy Observatories, which are operated by the Association of
Universities for Research in Astronomy, Inc., under cooperative
agreement with the National Science Foundation.}
Gemini GMOS package.  For the spectral data,
after initial processing with the GPREPARE task, all raw images
were reduced using
the GSREDUCE routine, which includes bias frame subtraction 
(created with the GBIAS task) and
division by normalized flat field frames (created with the GSFLAT task).
The CuAr arc lamp exposures were then used to derive the wavelength
calibration and rectify spectral images of 
each target and standard star.
After subtraction of telluric OH lines, 
spectra were optimally
extracted using the GSEXTRACT routine with standard settings.  
Target spectra were flux calibrated 
with the GSCALIBRATE routine and data for G 191-B2B.  
As telluric calibration sources were not included in this program,
telluric absorption features (e.g., O$_2$ and {\wat} bands) 
are present in the reduced spectral data.

\subsection{LDSS-3 Spectroscopy}

Optical spectra of SSSPM 1013-1356, 
2MASS 1227-0447 and 2MASS 1640+1231 were obtained
during 7-8 May 2006 (UT) using the Low Dispersion Survey Spectrograph (LDSS-3)
mounted on the Magellan 6.5m Clay Telescope (see Table~\ref{tab_log}).
LDSS-3 is an imaging spectrograph, upgraded by M.\ Gladders from the original
LDSS-2 \citep{all94} for improved
red sensitivity.  The instrument is composed of an
STA0500A 4K$\times$4K CCD camera that re-images an
8$\farcm$3 diameter field of view
at a pixel scale of 0$\farcs$189.  
For our observations, 
we employed the VPH-red grism (660 lines/mm) with a 0$\farcs$75
(4 pixels) wide longslit mask to
acquire 6050--10500 {\AA} spectra across the entire chip
with an average resolution {\ldl} $\approx$ 1800.  Dispersion
along the chip was 1.2 {\AA}/pixel.  The OG590 longpass filter
was used to eliminate second order light shortward of 6000 {\AA}.
Two slow-read exposures (to reduce read noise)
of 300 and 600~s each were obtained for 2MASS 1227-0447 
and 2MASS 1640+1231, respectively, while SSSPM~1013-1356 was observed
in two exposures of 300 and 450~s. We also obtained spectra of
nearby G2~V stars immediately after the subdwarf targets,
with differential airmasses ${\Delta}z < 0.1$,
for telluric absorption correction.  All spectral observations 
were accompanied by HeNeAr arc lamp and flat-field
quartz lamp exposures for dispersion and pixel response calibration.

LDSS-3 data were also reduced in the IRAF environment.
Raw science images
were first trimmed and subtracted by a median combined
set of slow-read bias frames taken during the afternoon.  
The resulting images were then divided by the corresponding
normalized, median-combined and bias-subtracted set of flat field frames.
The G star spectra were
extracted first using the APALL task, utilizing background subtraction
and optimal extraction options.  The subdwarf spectra were then optimally extracted using
the G star dispersion trace as a template.  Dispersion solutions were
determined from the arc lamp spectra extracted using the same dispersion trace; 
solutions were typically accurate
to 0.05--0.09 pixels, or 0.07--0.11 {\AA}.  
Flux calibration was determined
using the tasks STANDARD and SENSFUNC with
observations of the DA white dwarf spectral 
flux standard LTT 7987 
\citep[a.k.a.\ GJ 2147]{ham94}
obtained on 7 May 2006 (UT) with the same slit and grism combination
as the science data.  This calibration is adequate over the
spectral range 6000--10000 {\AA}. Corrections to telluric O$_2$ (6860--6960 {\AA} B-band,
7580--7700 {\AA} A-band)
and H$_2$O (7150--7300, 9270--9675 {\AA}) absorption bands
for each subdwarf/G star pair
were determined by linearly interpolating over these features in the 
G star spectrum, dividing by the uncorrected G star spectrum, and multiplying the
result with the subdwarf spectrum.  The two spectral exposures 
on each chip were combined after
median scaling, and the subsequent long and short wavelength data 
stitched together with no additional flux scaling.

Figure~\ref{fig_gmosvsldss3} shows a comparison between normalized 
GMOS and LDSS-3 spectral data for SSSPM~1013-1356.  Over most of the spectral 
range these spectra are roughly equivalent, 
with relative deviations (${\Delta}f_{\lambda}/f_{\lambda}$)
of less than 10\% for
$\lambda < 9000$~{\AA} (and generally less than 5\% for $\lambda < 8500$~{\AA}).
There is significant relative deviations between the spectra (10--50\%) in
the 9000--9900~{\AA} range, however, which we
attribute to unidentified flux calibration errors possibly due to the
uncorrected telluric absorption in the GMOS data 
(this region hosts the 9300~{\AA} stellar and telluric {\wat} bands).
We therefore focus our analysis on the 6000--9000~{\AA} region, 
which appears to be properly flux calibrated.

\section{Results}

\subsection{Characterizing the Spectra}

Reduced spectra for six of the targets (excluding 2MASS 1227-0447)
are displayed in Figure~\ref{fig_spectra}.  
These spectra exhibit a broad diversity of features and 
band strengths, but all are consistent with late-type dwarf
spectral morphologies.
Equivalent widths (EW) for the metal lines
observed in these spectra were measured using the IRAF SPLOT routine,
with uncertainties derived from multiple measurements on 
each of the individual spectra.
Values are listed in Table~\ref{tab_ews}.

2MASS 0851-0005 and 2MASS 0934+0523 have optical spectra
similar to late-type M dwarfs, with strong TiO 
(6400, 6650, 7050, 7750, 8450 and 8900 {\AA})
and VO (7400, 7900 and 8600 {\AA}) bands, 
CaH absorption at 6750~{\AA}, strong {\ki} (7665 and 7699 {\AA}) and
{\nai} (8183 and 8195 {\AA}) doublet lines, and red
spectral slopes from 7000 to 9000 {\AA}.  Weak {\wat} absorption is
also present at 9250 {\AA} in the spectra of both sources 
(it is blended with telluric
absorption beyond 9300 {\AA}), as is the Wing-Ford band 
of FeH at 9896 {\AA}. 

The spectra of 2MASS 1640+1231, 2MASS 0142+0523 and SSSPM 1013-1356
share many of these features, but also exhibit 
signatures of metal-deficient atmospheres.
Most notable are weakened bands of TiO and VO, many of
which are absent in the spectra of 2MASS 0142+0523
and SSSPM 1013-1356.  A key exception is the 7050 {\AA}
TiO band, which has a weakened blue wing 
but a persistent and prominent red bandhead. 
In contrast, CaH bands at 6350 and 6850 {\AA} are quite strong,
CrH and FeH bands at 8611 and 8692 {\AA} are clearly present,
and the 9896 {\AA} FeH band is considerably stronger than the M dwarfs.
The weakening of metal oxide bands has revealed
numerous metal lines, including
strong {\ki} and {\nai} lines (note the pressure-broadened
{\nai} D lines at 5890 and 5896 {\AA} in the spectra of 2MASS 0142+0523
and SSSPM 1013-1356); weak {\rbi} lines
at 7800 and 7948 {\AA}, {\cai} lines at 6120 and 6573 {\AA};
{\tii} at 8435 {\AA};
and a {\caii} triplet at 8498, 8542 and 8662 {\AA}.
Several of these lines have been previously identified
in the optical spectra of late-type sdMs and esdMs
\citep{lep1425,lep0822,melehpm2-59}. 
We also identify a new pair of {\tii} lines at 7209 and 7213 {\AA}
arising from the (4F)4p$\rightarrow$(4F)4s and (3P)sp$\rightarrow$(4P)4s transitions,
respectively \citep{kur88,mar88}.  
{\nai} and {\tii} lines have similar strengths between these three objects,
although {\rbi} lines are stronger
and {\cai} and {\caii} lines weaker in the spectrum of 2MASS 1640+1231.
The 8000--9000 {\AA} spectral continua of 2MASS 1640+1231,
2MASS 0142+0523 and SSSPM 1013-1356 are notably
less red than those of 2MASS 0851-0005 and 2MASS 0934+0536,
consistent with their photometric colors (Table~\ref{tab_targets}).

The spectrum of 2MASS~1626+3925, in turn, bears some resemblance to those
of 2MASS~0142+0523 and SSSPM~1013-1356, but is clearly of
later type.  The 7665/7699 {\AA} {\ki} doublet
has broadened considerably, mimicking pressure-broadening trends
observed between late-type M and L field dwarfs \citep{kir99,bur00}.  
The {\ki} wings suppress flux between 7300 and 8200 {\AA}, while
the weakening of features over 5800--6600 may be the result of
pressure-broadened {\nai} absorption (cf.~\citet{rei00}).
{\rbi} lines at 7800 and 7948 {\AA} are also 
more prominent, and {\csi} has appeared at 8521 {\AA}.
All three lines are similar in strength to those observed
in early- and mid-type L field dwarfs \citep{kir99}.
FeH, CrH and {\wat} bands are all significantly
stronger in the spectrum of this object,
and the 9896 {\AA} band is only surpassed
in strength by the L subdwarf 2MASS 0532+8246 \citep{me0532}.
On the other hand, the 8183/8195 {\AA} {\nai} doublet is 
weaker relative to the other spectra.  Since these lines arise
from excited lower energy states ($E_{lower}$ = 2.1 eV; \citet{kur95}), 
their weakening suggests
a lower photospheric temperature for 2MASS 1626+3925.
Lower temperature also explains the disappearance of the 
\ion{Ca}{2} triplet given the high ionization
potential of this element (6.11 eV).
The 8000--10000 {\AA} spectral slope of 2MASS 1626+3925 is notably redder
than those of 2MASS 1640+1231, 2MASS 0142+0523 and SSSPM 1013-1356,
and similar to those of 2MASS 0851-0005 and 2MASS 0934+0536.
TiO bands persist at 7050 and 8400 {\AA} but are much weaker
than those in the other spectra.  The observed spectral characteristics 
of 2MASS~1626+3925 are similar to those reported by \citet{giz07}. 

The spectrum of 2MASS 1227-0447 is shown in Figure~\ref{fig_esdm}, along
with the previously identified late-type esdMs SSSPM 0500-5406 
\citep[esdM6.5]{lod05} and LEHPM 2-59 \citep[esdM8]{melehpm2-59}.
These spectra show strong similarities, with extremely weak or absent
TiO and VO bands (compare the weak 7050 {\AA} band to the spectra in 
Figure~\ref{fig_spectra}) and strong CaH absorption.  Again, a forest of 
{\cai} and {\tii} lines are revealed in the spectrum of 2MASS 1227-0447
by the weakened metal oxide bands,
although they are somewhat weaker than those in 2MASS 1640+1231, 
2MASS 0142+0523 and SSSPM 1013-1356.  {\caii} lines, on the other hand,
are stronger, suggesting a relatively warm photosphere. The weak {\rbi}
lines that are present in the spectrum of this object are
similar in strength to LEHPM 2-59.  These two
sources also share similar {\ki} and {\nai} line strengths.
The overall spectral slope of this source
is somewhat redder than those of the other two esdMs shown in Figure~\ref{fig_esdm}, 
as discussed further in $\S$~4.4.

Finally, we note that 6563 {\AA} H$\alpha$ emission, an indicator of magnetic activity common in late-type M dwarf spectra, is seen weakly only
in the spectra of 2MASS 0851-0005 (EW = -1.0$\pm$0.5 {\AA}) and
2MASS 0934+0536 (EW = -1.5$\pm$0.5 {\AA}).  This indicates that 
our sources are 
magnetically weak or inactive, suggesting older ages \citep{haw99,rei03}.
The 6708 {\AA} {\lii} line, observed in absorption in the spectra of 
low mass M and L-type brown
dwarfs (M $<$ 0.065M$_{\sun}$; \citet{reb92}), is not present
in any of the spectra examined here.

\subsection{Radial Velocities}

Radial velocities for each source were measured using the detected metal lines
listed in Table~\ref{tab_ews}.
Line centers measured from Gaussian fits to the line cores for each individual spectrum
were compared 
to vacuum wavelengths listed in the Kurucz Atomic Line Database\footnote{Obtained
through the online database search form created by C.\ Heise and maintained
by P.\ Smith; see 
\url{http://cfa-www.harvard.edu/amdata/ampdata/kurucz23/sekur.html}.} 
\citep{kur95}.
The mean and standard deviations of these shifts (for all spectra and line features) 
are given in Table~\ref{tab_kinematics}, and include a systematic uncertainty of 5~{\kms} 
based on the typical scatter in arc lamp dispersion solutions.
For our GMOS observations of 2MASS 1626+3925 and SSSPM 1013-1356, we found significant 
differences in the derived radial velocities between the individual spectra, 
as large as 150~{\kms} for 2MASS 1626+3925, despite typical uncertainties of 10--13~{\kms} 
for each individual spectrum (based on the scatter between different
line measurements).  We attribute these discrepancies
to an unidentified systematic error in the wavelength calibration of the data.
Fortunately, SSSPM 1013-1356 was also observed with LDSS-3, for which
multiple spectral observations did not show such systematic effects, and
we utilize the derived radial velocity (50$\pm$7~{\kms}, statistically
consistent with the measurement of Scholz et al.\ 2004a) for our analysis.
For 2MASS 1626+3925, we make use of the radial velocity 
measurement of \citet{me1626}, $-$260$\pm$35~{\kms}, based on near-infrared spectroscopy.

All of the sources have relatively large
radial motions, as expected for halo and/or thick disk stars.
Perhaps most surprising are the motions of 2MASS 0851-0005
(72$\pm$24 {\kms}) and 2MASS 0934+0536 (156$\pm$28 {\kms}), 
both of which appear to have M field dwarf spectral 
characteristics.   The kinematics of our sample are discussed further in $\S$~4.6.

\section{Spectral Classification}

\subsection{Metallicity Groups}

Based on their spectral characteristics, our ultracool subdwarf
candidates appear to encompass the three main metallicity
classes defined by 
\citet{giz97}.
These qualitative assessments can be quantitatively
verified by comparing the spectral ratios
TiO5, CaH2 and CaH3
used by \citet{giz97} and \citet{lep03}
to define boundaries between the M dwarf metallicity groups.
Figure~\ref{fig_cah23tio}
shows the combined CaH2+CaH3 and TiO5 ratios for the sources
observed here and comparable measurements for late-type dwarfs, sdMs and esdMs
from the literature\footnote{See \citet{haw96,giz97,giz97b,lep03,lep1425,lep0822,sch1013,sch1444,rei05}; 
and \citet{melehpm2-59}.}.
The metallicity class divisions defined
by \citet{melehpm2-59} are overlain.
Both 2MASS 0851-0005 and 2MASS 0934+0536 lie within the dwarf locus
in this plot, with index values similar to M7-M9 dwarfs in the Palomar-MSU
sample \citep{haw96}.  2MASS 0142+0523 and SSSPM 1013-1356
lie well within the region occupied by
subdwarfs. 
2MASS 1227-0447 lies at the tail end of the extreme subdwarf locus, 
and its indices indicate both low metallicity (strong CaH and weak
TiO) and low {\teff}.
2MASS 1640+1231 and 2MASS 1626+3925 are difficult cases as their
CaH2+CaH3 indices lie below the limits of the \citet{melehpm2-59} dM/sdM
and sdM/esdM delineations.  The TiO5 ratio of 
2MASS 1640+1231 is the lowest reported to date, just slightly
below that of the high proper motion
source SSSPM 1444-2019 \citep{sch1444}, which also
has similar CaH2+CaH3 ratios and near-infrared colors (Figure~\ref{fig_cah23tio}). 
2MASS 1626+3925 has the lowest CaH2+CaH3 ratio reported to date.  
Both sources nevertheless appear to be very late-type subdwarfs
based on their spectral properties (Figure~\ref{fig_spectra}).

\subsection{M Dwarfs: 2MASS 0851-0005 and 2MASS 0934+0536}

Numerical subtypes for the two M dwarfs 2MASS 0851-0005 and 2MASS 0934+0536 were
initially determined using the spectral index scheme of
\citet{lep03}.  This method uses eight spectral indices 
from \citet{rei95}; \citet{haw02}; and their own work to measure
CaH, TiO and VO bandstrengths and spectral color
in the 6350--8500 {\AA} spectral band.
Three of the spectral type relations
based on the TiO5, CaH2 and CaH3 indices are identical to those  
used by \citet{giz97}, but are applicable only for spectral types M6 and earlier.
The remaining indices --- 
VO1, TiO6, VO2, TiO7 and Color-M --- are useable to spectral type $\sim$M9.
All of these indices were measured for the spectra after shifting
them to their rest frame velocities, and values are listed in  
Table~\ref{tab_class}.  Numerical subtypes for the two M dwarfs were assigned using
Eqns.~1--8 in \citet{lep03}.  The mean of
these individual numerical subtypes, rounded to the nearest half-type, yield
classifications of M7 for 2MASS 0934+0536 and M8 for 2MASS 0851-0005, albeit with
significant scatter (0.9 subtypes) in the latter case.

To ascertain the robustness of these classifications, in Figure~\ref{fig_mdwcomp} we 
compare normalized optical spectra for the two sources
to the spectral standards VB~8 (M7) and VB~10 (M8)
from \citet{cru03}\footnote{The \citet{cru03} data were
acquired with the RC Spectrograph mounted on the Kitt Peak National Observatory
4m telescope; and reduced in a similar manner as the GMOS and LDSS-3
data.  We therefore assume that this data are equivalently flux
calibrated over the 6000--9000~{\AA} range.}, as shown
in Figure~\ref{fig_mdwcomp}.
2MASS 0934+0536 is an excellent match to the M7 standard
over the 6200--9000 {\AA} region.
The spectrum of 2MASS 0851-0005, on the other hand, 
shows clear deviations from that of VB~10, with
an overall redder spectral energy
distribution, weaker TiO and VO absorption at 7800--8000 {\AA}
and stronger CaH absorption at 7000 {\AA}.  Interestingly, some of the
same spectral deviations are seen in another high velocity late-type M dwarf,
LSR 1826+3014 \citep{lep1826}, an object with halo kinematics but
dwarf-like spectral features. The redder spectral energy distribution
of this source provides a better match to the spectrum of 2MASS 0851-0005, 
albeit with somewhat stronger VO absorption at 7400 {\AA}.
As both LSR 1826+3014 and 2MASS 0851-0005 both exhibit
large motions, these subtle spectral differences may be indicative of 
either older age (high surface gravity) or slight metal deficiency.

\subsection{M Subdwarfs: 2MASS 0142+0523, SSSPM 1013-1356 and 2MASS 1640+1231}

2MASS 0142+0523, SSSPM 1013-1356 and 2MASS 1640+1231 were 
classified using the CaH2 and CaH3 index/spectral type relations of 
\citet[also employed by \citet{lep03}]{giz97}.
These relations were established only for spectral types as late as sdM7,
but we follow current practice (e.g., Scholz et al.\ 2004a,b) in
extrapolating them to later subtypes (see $\S$~6.1.1).
Averaging the CaH subtypes yields classifications of
sdM8.5 for 2MASS 0142+0523, sdM9 for 2MASS 1640+1231 and
sdM9.5 for SSSPM 1013-1356 (equivalent for both the GMOS and LDSS-3 data
for this source). Our classification of SSSPM 1013-1356 is identical to that derived
by \citet{sch1013}.  \citet{lep03} defined a third index/spectral type
relation for M subdwarfs using the Color-M index; the associated subtypes for this
index are listed in Table~\ref{tab_class} but were not used to determine the 
classifications of these sources.

The classifications of 2MASS 0142+0523, 2MASS 1640+1231 and SSSPM 1013-1356
suggest a natural sequence between these three objects; however,
inspection of the spectra in Figure~\ref{fig_spectra}
shows that this is not the case.  2MASS 1640+1231 clearly exhibits
stronger TiO absorption at 7800 and 8500 {\AA} than the other two 
objects, while the 6600 {\AA} bump between the 6400 and 6800 {\AA} CaH bands
is less pronounced.  {\caii} lines are also weaker, and
2MASS 1640+1231 exhibits a redder spectral
slope from 6500--8000 {\AA}.  We interpret these discrepancies
as the result of a higher metallicity in this source as compared
to both 2MASS 0142+0523 and SSSPM 1013-1356.  Higher metallicities should produce
features more consistent with M field dwarfs,
and the spectral discrepancies in 2MASS 1640+1231 
are indeed those same features that are prominent in the spectra
of 2MASS 0851-0005 and 2MASS 0934+0536.  
This is demonstrated in Figure~\ref{fig_zcomp}, which compares
the normalized spectra of 2MASS 0851-0005, 2MASS 1640+1231 and SSSPM 1013-1356.
Differences between these spectra can
be readily attributed to the weakening of TiO bands in this sequence,
and the onset and strengthening of CaH, FeH and CrH bands.
Given the similarity in {\ki} and {\nai} line strengths, these 
variations likely arise
from differences in metallicity (cf.\ Fig.~2 in \citet{lep1425}).
2MASS 1640+1231 is a clear intermediary 
between the M8 2MASS 0851-0005 and the sdM9.5 SSSPM 1013-1356,
yet is distinct from both sources. 
As such, we assign a classification of d/sdM9 for this source,
the prefix denoting it as a ``mild subdwarf'' (cf., \citet{mou78}).
This designation is discussed in further detail in $\S$~6.1.2.

We noted previously that the TiO5, CaH2 and CaH3 indices and near-infrared
colors of 2MASS 1640+1231 were similar to those of the 
high proper motion star SSSPM 1444-2019.  
Figure~\ref{fig_1640comp} demonstrates that their optical spectra are 
also similar, as also reported in \citet{giz07}.  
Both exhibit molecular bands that are intermediate
in strength between M dwarfs and subdwarfs, but with equivalent
alkali lines.  SSSPM~1444+2019 may be slightly 
less metal-poor than 2MASS~1640+1231, as it exhibits 
a fainter flux peak
between the 7050~{\AA} TiO band and {\ki} doublet, and more absorption
in the region of the 7750~{\AA} TiO and 7900~{\AA} VO bands.
These differences are subtle, however, and we tentatively 
propose that both sources be classified as mild subdwarfs, d/sdM9,
to reflect their apparent intermediate metallicities 
(see also Scholz et al.\ 2004, Figure~4).

\subsection{M Extreme Subdwarf: 2MASS 1227-0447}

For 2MASS 1227-0447, we employed the esdM CaH2 and CaH3 ratio/spectral type relations
of \citet{giz97}, which also must be extrapolated beyond their 
originally defined
spectral type ranges.  These indices yield a mean subtype of sdM7.5.  
Figure~\ref{fig_esdm} demonstrates that this classification is consistent
with the spectral morphology of this source as compared to the
two late-type esdMs SSSPM 0500-5406 (esdM6.5) and LEHPM 2-59 (esdM8); 
in particular, the intermediate {\ki} and
{\nai} line strengths and the presence of {\rbi} lines.  2MASS~1227-0447 is
only the third ultracool extreme subdwarf identified to date 
\citep{sch99,melehpm2-59}

As mentioned above, the 6500--8000~{\AA} spectral slope of 2MASS 1227-0447
is somewhat redder than those of SSSPM 0500-5406 and LEHPM 2-59,
consistent with its redder $R_{ESO}-I_N$ colors (1.9 versus 1.6 for 
both of the latter; SSS).  As a result, the Color-M index/spectral type relation
of \citet{lep03} gives a very late, and inconsistent, type
of esdL5 (or esdM15). 
The origin of this red spectral slope is unclear.  
It could be extrinsic, due to reddening from interstellar dust grains,
although this source does not appear to be particularly distant ($\S$~4.6).
Alternately, 2MASS 1227-0447 
may be less metal-poor than the other esdMs,
resulting in a spectral slope more similar to late-type sdMs;
or cooler, resulting in more flux arising at longer wavelengths.
However, spectral model fits discussed in $\S$~5 do not provide
support for either of these scenarios.  Additional 
observations are required to ascertain the nature of the red spectral slope
exhibited by this late-type esdM.

\subsection{L Subdwarf: 2MASS 1626+3925}

2MASS 1626+3925 exhibits features
most similar to L field dwarfs; hence, use of the \citet{giz97}
and \citet{lep03} M subdwarf classification schemes is
inappropriate.  On the other hand, the L dwarf
classification scheme of \citet{kir99} is similarly problematic,
as it is tied to features (TiO, VO, CrH bands and spectral color) that
are known to be sensitive to metallicity.
How then to assign a subtype to this object?  We chose the tactic
of directly comparing L dwarf spectral standards from \citet{kir99}
to the spectrum of 2MASS 1626+3925 and 
selecting the subtype that provided the closest overall match.
Figure~\ref{fig_2m1626comp} illustrates that best match, to the L4
standard 2MASS J11550087+2307058 (hereafter 2MASS~1155+2307;
see also \citet{giz07}). 
There is reasonable agreement between the spectra of these
sources over the 7300--9000 {\AA} range, particularly around
the pressure-broadened {\ki} doublet, {\rbi} and {\csi} lines,
the 8611 and 8692 {\AA} CrH and FeH
bands, and the overall spectral slope in this waveband.  
The 8183/8195 {\AA} {\nai} lines
are notably stronger in the spectrum of 2MASS 1626+3925. 
As these lines arise from excited energy states, their 
strength suggests that the photosphere of
2MASS 1626+3925 may be warmer than that of 2MASS 1155+2307.
TiO bands at 7050 and 8400 {\AA} and CaH bands 
at 6400 and 6800 {\AA} are also stronger in the spectrum
of this source. 
We attribute these discrepancies to metallicity effects.
In the absence of a robust classification scheme for L
subdwarfs, we assign a type of sdL4 to 2MASS 1626+3925 based on its
similarity to 2MASS 1155+2307.  L subdwarf classification
is discussed in further detail in $\S$~6.1.3.

\subsection{Spectrophotometric Distance Estimates and Kinematics}

Spectrophotometric distance estimates for the sources in our sample
were determined using established absolute magnitude/spectral type relations
and the observed apparent magnitudes.  Results are listed in Table~\ref{tab_kinematics}.
For the M dwarfs 2MASS 0851-0005 and 2MASS 0934+0536, we averaged estimates 
from the $M_J$ and $M_K$/spectral type relations from
\citet{dah02,cru03}; and \citet{rei05}.  Both sources,
which are intrinsically faint, lie nearly 100 pc from the Sun.
For the subdwarfs, because
few late-type sdMs and esdMs currently have measured parallaxes\footnote{The 
latest-type M subdwarfs with parallaxes are the sdM7 LHS 377 and the esdM6 LHS 1742a
\citep{mon92}, both of which have earlier spectral types than the sources
examined here.}, we
extrapolated the $M_R$ and $M_K$/spectral type relations of
\citet{lep03} and \citet{rei05}.  Our distance estimate for SSSPM 1013-1356,
$\sim$30 pc, is lower than but not inconsistent with that of \citet[50$\pm$15~pc]{sch1013}.
For 2MASS 1640+1231, we combined
both dwarf and subdwarf distance estimates
which differ by roughly 20\%.  Extrapolating these relations for the 
L subdwarf 2MASS 1626+3925 is likely to be too extreme, particularly with its
somewhat ad hoc 
spectral classification, so a distance estimate was made using
the the $M_J$/spectral type relations 
for L dwarfs from \citet{dah02,cru03}; and \citet{vrb04} assuming a type L4.
The relatively nearby estimated distance for this source, $\sim$20~pc, should be
considered with caution; however, it does make 2MASS 1626+3925 
a high priority target for parallax observations.

Using these estimated distances and the measured proper motions
and radial velocities, we computed
$UVW$ space velocity components relative to the Local Standard of Rest (LSR)
for the six sources in our sample with measured proper motions (i.e., excluding
2MASS 0934+0536).  The LSR solar motion was assumed to be
[$U_{\sun},V_{\sun},W_{\sun}$] = [10,5,7] {\kms} \citep{deh98}.  
Figure~\ref{fig_uvw} illustrates these velocities relative to
the 3$\sigma$ velocity dispersion sphere of local disk M dwarfs \citep{haw96},
thick disk stars \citep{str87} and halo stars \citep{som90}.
The M dwarf 2MASS 0851-0005 is the only source encompassed within
the disk dwarf velocity sphere; the remaining sources have motions
consistent with either thick disk or halo populations.
2MASS 1227-0447, 2MASS 1626+3925 and 2MASS 1640+1231 are particularly
good candidate halo objects.  These kinematics, while not capable of
conclusively determining population membership for any one
source, are nonetheless consistent with expectations 
based on their metallicity classifications.  The most
metal-poor (metal-rich) objects are associated with the oldest 
(youngest) populations.

\section{Spectral Model Fits}

To further gauge the physical properties of these sources, we 
compared the spectral data to subsolar metallicity
spectral models from \citet[NextGen]{hau99}
and \citet[Cond]{all01}.  Details on the 
characteristics and differences between these model sets are 
described more fully in \citet{all01} and references therein.
The NextGen models and their antecedents 
have been used previously for
fitting M subdwarf spectra \citep{giz97,sch99,daw00,leg00,lep0822,melehpm2-59}.

The model fitting procedure followed here is similar
to that described in \citet{melehpm2-59}.  
We sampled grids of both model sets
spanning temperatures of 2200 $\leq$ \teff $\leq$ 3200~K
in steps of 100~K, metallicities of
-3.0 $\leq$ [M/H] $\leq$ 0.0~dex in steps of 0.5~dex, 
and a fixed surface gravity of
$\log{g}$ = 5.5 {\cmss}.  
Both empirical and model spectra
were normalized at 8100~{\AA}, and the observed
data were shifted to their rest frame velocities.
Model spectra were also reduced in resolution to that of
the observed data using a Gaussian convolution kernel.
For each spectrum/model pairing,
the chi-square deviation ($\chi^2$ = $\sum[f_{\lambda}^{\rm Observed}-f_{\lambda}^{\rm Model}]^2/f_{\lambda}^{\rm Model}$)
was computed over the spectral
range 6200--9000 {\AA}. The normalization of the 
model spectra was allowed to vary slightly for continuum offsets, and the normalization
with the minimum $\chi^2$ was retained.  For each
source and model combination, we computed the best fitting
{\teff} and [M/H] values by averaging the parameters for the
eight best model fits weighted by their associated $\chi^2$ values.
These values are listed in Table~\ref{tab_modelfits}; reported
uncertainties take into account the range of best-fit parameters
but not systematic effects (see below)

Figures~\ref{fig_modelfits} and~\ref{fig_modelfits2} illustrate
the quality of the spectral model fits.
For the M dwarfs 2MASS 0851-0005 and 2MASS 0934+0536, 
the NextGen models provide a somewhat better fit shortward
of 8300 {\AA}, albeit with a poorer match to the 6400--6600 {\AA} bump
between the CaH bands, and the bandhead of the 7050 {\AA} TiO band
(which is poorly reproduced in all of the models). There is excess absorption
at 8400 {\AA}, likely due to errors in the TiO bandstrengths at this wavelength.
Likewise, the COND models exhibit stronger VO absorption at 8600 {\AA} than is 
observed.  The overall spectral shapes match fairly well; however,
for the M subdwarfs 2MASS 0142+0523 and SSSPM 1013-1356, 
the COND models provide slightly better fits, particularly shortward of 7000 {\AA} 
and redward of the {\ki} lines. 
Note that the 7300--7500 {\AA} region is poorly matched by both sets of models;
this is likely due to the absence of the 7400 {\AA} VO band
in the model opacity set (P.\ Hauschildt, 2006, priv.\ comm.).   The NextGen models do a better
job at matching the broad and deep 6800 {\AA} CaH band seen in the subdwarfs, 
as is particularly
evident in fits to the esdM7.5 2MASS 1227-0447.  
For this object, NextGen models provide a better match shortward of 7200 {\AA},
while COND models fit the pseudocontinuum longward of the {\ki} lines.
Neither model matches the spectrum of 
2MASS 1626+3925 adequately.  
While there is agreement in critical absorption regions
(6800 {\AA} CaH bands, {\ki} lines, and 7600--7800 {\AA} continuum),
there are also large discrepancies in the flux peaks between these features.  

While the models do not provide accurate fits to the data, implying the
likely presence of systematic biases in
derived parameters, it is nevertheless worthwhile to examine trends in these
parameters.
Overall, metallicities scale well with the assigned classifications, 
decreasing between the M dwarfs ([M/H] $\sim$ -0.2),
sdMs ([M/H] $\sim$ -0.9) and the single esdM ([M/H] $\sim$ -1.4).
The derived metallicity of 2MASS 1640+1231, intermediate between the
dwarfs and subdwarfs, is also consistent with its
mild subdwarf classification.
The best fitting spectral models suggest a very low metallicity
for 2MASS 1626+3925, although the poor quality of the 
fits encourages skepticism in the derived values.
Within the dwarf and subdwarf metallicity classes, 
there is also a trend of lower {\teff}s with later subtypes.
2MASS 1227-0447 has a similar {\teff} as the esdM8 LEHPM 2-59
and is 50-100~K cooler than the esdM7 APMPM 0559-2903 \citep{melehpm2-59}, 
consistent with its intermediate spectral type.
Across metallicity classes, however, there appears to be less consistency
between numerical types and {\teff}.  2MASS 1227-0447
is $\sim$300~K hotter than the M dwarfs 
2MASS 0934+0536 and 2MASS 0851-0005, despite having a similar
numerical type.  2MASS 0142+0523 and SSSPM 1013-1356
are 100-200~K hotter than 2MASS 0851-0005, despite having later
numerical types.  Even 2MASS 1626+3925 has a {\teff} that
is $\sim$600~K hotter
than the typical L4 field dwarf \citep{vrb04}, although this comparison
is questionable.
The overall trend --- hotter {\teff}s for more
metal-poor objects at a given numerical classification --- has been
previously noted by \citet{melehpm2-59} between ultracool
dwarfs and extreme subdwarfs, and attributed to the classification
methodology.  Our results indicate
that similar {\teff} offsets may occur for all ultracool
metallicity classes.

\section{Discussion}

\subsection{Cool and Ultracool Subdwarf Classification}

\subsubsection{The Need for a Revised Classification Scheme}

With the exceptions of the M dwarfs 2MASS 0851-0005 and 2MASS 0934+0536,
all of the sources examined here --- and indeed all of the ultracool
subdwarfs currently known --- have been classified using either 
extrapolations of existing classification schemes \citep{giz97,lep03}
or direct comparisons to sources with very
different metallicities (e.g., 2MASS 1626+3925).  It is clear that existing
schemes are wholly inadequate for the new generation of ultracool
metal-poor objects for several reasons.  First, the use of the TiO5 
index as a metallicity indicator
becomes increasingly problematic for later spectral types as this
band begins to weaken either through depletion onto condensate dust grains (see $\S$~6.2)
or increasing absorption from proximate species such as pressure-broadened
{\nai} and {\ki} lines and CaH absorption.  This has already been recognized in
field dwarfs \citep{giz97} and L subdwarfs \citep{lep1610,me0532}, but may also
be a problem for the latest-type M subdwarfs.
Second, the 6300 and 6800 {\AA} CaH bands
which are the focus of the \citet{giz97} scheme become largely saturated
in the latest-type subdwarfs, leading to relatively small differences in 
spectral morphology between subtypes.   Two examples illustrate this point.
2MASS 0142+0523 and SSSPM 1013-1356 are separated by a full subclass
based on extrapolations of the \citet{giz97} index/spectral type
relations, but differences between the spectra are relatively subtle
(Figure~\ref{fig_latesd}).  The esdMs SSSPM~0500-5406
and LEHPM 2-59 in Figure~\ref{fig_esdm}
also show strong similarities, despite being separated by 1.5 subclasses.  
Finally, the spectral region encompassing these bands becomes increasingly
faint for later spectral types of all metallicity classes, leading to
classification errors due to low signal-to-noise data.
All of these effects minimize the utility of the spectral classifications
as estimators of absolute brightness, {\teff} and metallicity.

Based on studies of 
ultracool field dwarfs (e.g., \citet{kir91,kir99,lep03}), it
is clear that the spectral classification of cool and 
ultracool subdwarfs needs to shift to the redder
spectral bands and encompass a broader set of spectral features.
\citet{lep03} have made a first step in this direction through the
use of the Color-M index; however, as exemplified by the case of
2MASS 1227-0447, this pseudo-continuum index
can be overly sensitive to reddening. 
As a guide to what features may be useful in the ultracool
subdwarf regime, we show the spectra of five of the latest-type subdwarfs
currently known in Figure~\ref{fig_latesd}:
the sdM8 LSR 1425+7102 (data from \citet{lep1425}), the sdM8.5 2MASS 0142+0523,
the sdM9.5 SSSPM 1013-1356, the sdL4 2MASS 1626+3925 
and the sdL7 2MASS 0532+8246 (data from \citet{me0532}).
The features in these spectra show a natural progression, with the 
emergence and strengthening of {\rbi} lines at sdM8.5 and {\csi}
in the L subdwarfs; the decline of the 8183/8195 {\AA} {\nai} doublet;
substantial pressure broadening of the {\ki} doublet,
contributing in part to the reddening of the 7700-8600 {\AA} spectral slope;
weakening of the 6500--6700 {\AA} spectral bump and 7050 {\AA} TiO band,
possibly driven by both {\nai} and {\ki} pressure broadening; 
and the dramatic
strengthening of FeH and CrH bands at 8600-8700 and 9900~{\AA}.  
While these objects may span a range of both metallicities and {\teff}s
(Table~\ref{tab_modelfits}; see also Scholz et al.\ 2004b),
their spectra nevertheless form a fairly coherent sequence.

A more fundamental deficiency in the classification of subdwarfs is the lack
of any predefined spectral standards, one of the fundamental requirements for
a spectral classification scheme under the MK System \citep{mor43,mor73,kee76,cor94}.
For ultracool subdwarfs, the primary obstacle is the paucity of known
examples.  A sufficiently large number of objects is needed
to encompass the range of spectral morphologies as well as weed out
peculiar objects (such as LSR 1610-0040; Cushing \& Vacca 2006).
For early- and mid-type M subdwarfs, there are a plethora of well-studied
sources from which standards should be
identified before extending any classification 
scheme into the ultracool regime.  

\subsubsection{Mild Subdwarfs}

The demarcation of a mild subdwarf metallicity class to distinguish
SSSPM 1444-2019 and 2MASS 1640+1231 reflects recent 
discoveries of ultracool dwarfs that exhibit signs of slight metal 
deficiency.  Several groups have begun to uncover ``blue'' L dwarfs
\citep[K.\ L.\ Cruz, 2006, in prep.]{cru03,mewide3,kna04,chi06}, many
with relatively large proper motions suggesting thick disk or halo
kinematics.  One peculiar T dwarf, 2MASS J09373487+2931409, has
been shown to be slightly metal-poor ([M/H] $\sim$ -0.4 to -0.1)
on the basis of spectral model comparisons \citep{me03opt,metgrav}.

That slightly metal-poor ultracool dwarfs exist 
should not be surprising, as it is well known that hotter dwarf 
stars exhibit a range of both
metallicities and elemental abundance patterns.  However, spectral
signatures of slight metallicity differences are generally subtle in
earlier-type stars.
At lower {\teff}s, the presence of strong and overlapping
molecular bands results in greater sensitivity to metallicity. 
As a result, so-called
mild ultracool subdwarfs are more spectrally distinct from their
dwarf and subdwarf counterparts, and should be acknowledged as such.
We therefore advocate the use of the d/sd designation for these
slightly metal-poor stars and brown dwarfs.

\subsubsection{Defining the L Subdwarf Class}

We have classified 2MASS 1626+3925 as an sdL4 based on its similarity
to the L4 dwarf spectral standard over the 7300--9000 {\AA}
band.  By analogy, we propose a subtype
of sdL7 for 2MASS 0532+8246, given its similarity to the L7 field dwarf 
DENIS 0205-1159 \citep{del97,me0532}.  At this stage, this approach 
is the closest we can come to the determination of spectral 
types by comparison to
spectral standards.  It is a logical method, as it provides a direct
tie between the classification
of L subdwarfs to the well-defined scheme for L dwarfs, and should facilitate
comparisons between metallicity classes.  Ultimately, spectral
standards in the L subdwarf regime should be identified once a sufficiently
large number of such objects are discovered.

There is a broader issue of concern in the L subdwarf regime, however;
faced with uncertainty in the classification of low temperature
subdwarfs, how do we robustly
distinguish the threshold between M and L subdwarf classes?  
Based on the spectral properties
of the subdwarfs shown in Figure~\ref{fig_latesd} and analogous features
in L field dwarfs, we propose the following criteria:
\begin{itemize}
\item flattening of the 7400 {\AA} peak between the 7050 {\AA} TiO band and 
7700~{\AA} {\ki} doublet, 
and a turnover from a red slope to a blue slope in this spectral region;
\item saturation of the 7700 {\AA} {\ki} doublet;
\item decline in the 8183/8195 {\AA} {\nai} doublet;
\item appearance of the 8521 {\AA} {\csi} line; and
\item disappearance of the {\caii} triplet near 8500 {\AA}.
\end{itemize}
By these criteria, LSR 1610-0040 \citep{lep1610} does not meet the definition
of an L subdwarf, as its spectrum exhibits a red-sloped 7400 {\AA} peak,
unsaturated {\ki} lines and no {\csi} line.  Indeed, the strength of its 
{\ki} lines and somewhat weaker CaH, CrH and FeH bands suggests
that is rather a slightly metal-poor, late-type mild M subdwarf.
This may explain
in part its unusual near-infrared spectral properties \citep{cus06}
and relatively blue $I_N-J$ and red $J-K_S$ colors.
However, we stress that the L subdwarf criteria listed above are
preliminary suggestions, and a robust set of differentiators awaits
the discovery of more L subdwarfs.

\subsection{Dust Formation in Low-Temperature Metal-Poor Atmospheres}

One of the defining properties of L field dwarfs is the formation
of condensate dust in their photospheres, leading to a depletion
of gaseous TiO and VO bands and a general reddening of their
spectral energy distributions \citep{tsu96a,tsu96b,bur99,ack01,all01}.  
However, \citet{me0532} have
pointed out that the L subdwarf 2MASS 0532+8246, which 
has an optical spectrum similar to a typically dusty
late-type L dwarf, exhibits a distinct band of gaseous TiO at 8450 {\AA}, 
a species which should be weakened by the lower metallicity of this
object and depleted as Ti is sequestered onto dust grains. 
With 2MASS 1626+3925,
we see further evidence of a reduction in condensate formation.
Three bands of TiO are evident in the spectrum
of this source at 6650, 7050 and 8450 {\AA}, as are {\tii} lines.
Gaseous atomic calcium, which is removed 
from the photospheres of L dwarfs by various
Ca-Ti and Ca-Al condensates \citep{lod02},
is also present in the spectrum of 2MASS 1626+3925
in the form of CaH and {\cai}.  The persistence of these species
in gaseous form, in addition to the relatively blue near-infrared color
of this source, indicates that condensates have not formed in 
significant quantities in the atmosphere of this source.  
That two L subdwarfs exhibit similar peculiarities
suggests that dust formation in general
may be inhibited in these objects, perhaps due to a lack of sufficient
metals to drive grain formation.   One may also argue that the
apparently higher {\teff}s of metal-poor 
dwarfs as compared to equivalently classified
field dwarfs suggests that their atmospheres may not be cool enough
to form condensates in the first place.  However, chemical models by
\citet{lod02} have shown that at the higher pressures of subdwarf 
photospheres, Ti is incorporated into calcium titanates at higher 
temperatures, indicating the condensate formation can occur
at hotter {\teff}s in metal-poor objects.
Additional chemical modeling is clearly needed.

We note in passing that 2MASS 1626+3925 is somewhat bluer
in the near-infrared that the apparently later-type 
L subdwarf 2MASS 0532+8246 ($J-K_s$ = -0.03 versus 0.26).  This suggests
that some condensate formation may have occurred in the latter object,
or that it is less metal poor (cf.~\citet{sch1444}).  If the former
scenario is correct, the sensitivity of the condensate formation process on 
temperature, pressure and the order of condensate formation
\citep{bur99,lod02} implies that any condensate clouds present in the
atmosphere of this object may have very different compositions than those
in L field dwarf atmospheres.

\subsection{A Compendium of Known Ultracool Subdwarfs}

Table~\ref{tab_compendium}
provides a list of known
ultracool subdwarfs with optical spectra reported in the 
literature. A total of 16 such objects are now known, 
encompassing 4 M- and T-type mild subdwarfs (including LSR 1610-0040),
6 sdMs, 3 sdLs, and 3 esdMs.  We do not include peculiar M dwarfs
such as 2MASS~0851-0005 and LSR 1826+3014 as effects other than metallicity
may explain their fairly subtle spectral deviations.  
An overview of this list reveals several salient characteristics of ultracool subdwarfs.
Most are relatively faint, particularly in the optical (three are undetected
in $R$-band photographic plates), and lie at large distances (most are further than 50 pc),
most likely reflecting their low space density (e.g.~\citet{dig03}).
All have large proper motions, a possible bias due to their predominant
discovery in proper motion surveys; and large radial velocities, 
with nearly half exceeding 100 {\kms} of line-of-sight motion. 
Most of the sources have relatively blue $R-J$ and $J-K_s$ colors as compared to 
M and L field dwarfs, and esdMs are generally bluer than sdMs for a
given numerical type, as expected from the reduction of optical TiO and VO bandstrengths 
and enhanced H$_2$ absorption at 2 $\micron$. 
Finally, the sources are well distributed across the sky, as would be expected for
a low density, widely dispersed population of thick disk and/or halo stars.
With the recent inception of near-infrared proper motion surveys
(e.g.~\citet{dea05,art06}; J.\ D.\ Kirkpatrick, 2007, in prep.) which sample
the spectral bands in which very cool subdwarfs are brightest, it is likely
that this list will expand substantially in the near term.

\section{Summary}

We have presented optical spectra for a sample of seven ultracool subdwarf
candidates, identified in a color-selected search for T dwarfs using
the 2MASS All Sky Survey.  2MASS 0851-0005
and 2MASS 0934+0536 are distant ($\sim$100 pc),
late-type field M dwarfs, although the former 
exhibits spectral features suggesting an older age or slight metal deficiency.
2MASS 0142+0523 is a new sdM8.5, joining the previously identified sdM9.5
SSSPM 1013-1356 and sdL4 2MASS 1626+3925 as one of the latest-type subdwarfs known.
2MASS 1640+1231 is an example of a subclass of ultracool mild subdwarfs,
with metallicities intermediate between dwarfs and subdwarfs, and is similar to
the previously identified high proper motion star SSSPM 1444-2019.
2MASS 1227-0447 is a new ultracool extreme subdwarf, classified sdM7.5.  These
sources join a rapidly growing list of low mass, metal-poor dwarfs
whose properties are only beginning to be understood.
Much basic investigative work remains for these objects,
including the measurement of parallaxes and bolometric magnitudes,
the development of accurate spectral models to constrain physical properties
and investigate processes such as condensate formation,
and the construction of a robust spectral classification scheme.
As wide-field red optical photographic plate surveys have revealed a large population
of cool subdwarfs, the recent advent of wide-field near-infrared proper motions surveys
should greatly expand
our sampling of the ultracool subdwarf population and its properties.

\acknowledgments

We thank P.\ Hauschildt and F.\ Allard for making their
spectral models electronically available; S.\ L\'epine and
R.-D.\ Schultz for providing electronic versions of their
spectral data for LSR 1425+7102, LSR 1826+3014 and SSSPM 1444-2019;
and R.\ Cutri for providing mean 2MASS colors of various spectral types.
AJB acknowledges the assistance of Marcel Bergman, 
Tom Geballe and Inger Jorgensen during the planning and acquisition
of Gemini GMOS queue data; and Mauricio Hernandez and Hernan Nu\~{n}ez 
for their help with the Magellan observations.
We also thank our anonymous referee for her/his useful critique of the original
manuscript.
KLC is supported by an NSF Astronomy and Astrophysics Postdoctoral Fellowship under award AST-0401418.
Based on observations obtained at the Gemini Observatory, which is operated by the
Association of Universities for Research in Astronomy, Inc., under a cooperative agreement
with the NSF on behalf of the Gemini partnership: the National Science Foundation (United
States), the Particle Physics and Astronomy Research Council (United Kingdom), the
National Research Council (Canada), CONICYT (Chile), the Australian Research Council
(Australia), CNPq (Brazil) and CONICET (Argentina);
and the 6.5 meter 
Magellan Telescopes located at Las Campanas Observatory, Chile.
This publication makes use of data 
from the Two Micron All Sky Survey, which is a
joint project of the University of Massachusetts and the Infrared
Processing and Analysis Center, and funded by the National
Aeronautics and Space Administration and the National Science
Foundation.
2MASS data were obtained from the NASA/IPAC Infrared
Science Archive, which is operated by the Jet Propulsion
Laboratory, California Institute of Technology, under contract
with the National Aeronautics and Space Administration.

Facilities: \facility{Gemini Gillett(GMOS)}, \facility{Magellan Clay(LDSS-3)}

\clearpage

\begin{deluxetable}{lcccccccl}
\tabletypesize{\footnotesize}
\tablecaption{Ultracool Subdwarf Candidates. \label{tab_targets}}
\tablewidth{0pt}
\tablehead{
\colhead{Source\tablenotemark{a}} &
\colhead{$R_{ESO}$} &
\colhead{$I_N$} &
\colhead{$J$\tablenotemark{c}} &
\colhead{$J-H$\tablenotemark{c}} &
\colhead{$H-K_s$\tablenotemark{c}} &
\colhead{$\mu$\tablenotemark{d}} &
\colhead{$\theta$\tablenotemark{d}} &
\colhead{Ref.\tablenotemark{e}} \\
\colhead{} &
\colhead{} &
\colhead{} &
\colhead{} &
\colhead{} &
\colhead{} &
\colhead{($\arcsec$ yr$^{-1}$)} &
\colhead{($\degr$)} &
\colhead{} \\
}
\startdata
2MASS J01423153+0523285 & 19.49 & \nodata & 15.91$\pm$0.08 & 0.32$\pm$0.14 & -0.01$\pm$0.26 & 0$\farcs$63$\pm$0$\farcs$05 & 117$\pm$4 & 1 \\
2MASS J08514509-0005360 & \nodata & 18.65 & 15.99$\pm$0.09 & 0.70$\pm$0.12 & -0.14$\pm$0.25 & 0$\farcs$14$\pm$0$\farcs$05 & 206$\pm$20 &  2  \\
2MASS J09340617+0536234 & \nodata & 17.86 & 15.57$\pm$0.06 & 0.43$\pm$0.09 & 0.38$\pm$0.14 &  \nodata & \nodata & 2  \\
2MASS J10130734-1356204 & 18.73 & 16.01 & 14.62$\pm$0.03 & 0.24$\pm$0.06 & -0.02$\pm$0.09 & 1$\farcs$03$\pm$0$\farcs$10 & 176$\pm$5 &  3 \\
2MASS J12270506-0447207 & 18.48 & 16.63 & 15.49$\pm$0.05 & 0.28$\pm$0.08 & 0.33$\pm$0.13 & 0$\farcs$48$\pm$0$\farcs$02 & 59.7$\pm$0.2 &  2 \\
2MASS J16262034+3925190 & 19.67 & 16.68 & 14.44$\pm$0.03 & -0.10$\pm$0.06 & 0.07$\pm$0.09 & 1$\farcs$37$\pm$0$\farcs$10 & 280$\pm$4 & 4 \\
2MASS J16403197+1231068 & 20.63 & \nodata & 15.95$\pm$0.08 & 0.34$\pm$0.14 & $<$0.09 & 0$\farcs$82$\pm$0$\farcs$06 & 246$\pm$4 &  1 \\
\enddata
\tablenotetext{a}{All objects are listed with their 2MASS All Sky Data Release
source designations, given as ÔÔ2MASS Jhhmmss[.]ss$\pm$ddmmss[.]sÕÕ. The suffix is 
the sexagesimal right ascension and declination at J2000.0 equinox.}
\tablenotetext{b}{POSS-II and UKST photographic plate photometry from SuperCOSMOS Sky Survey \citep{ham01a,ham01b,ham01c}.}
\tablenotetext{c}{Near-infrared photometry from 2MASS \citep{skr06}.}
\tablenotetext{d}{Calculated from 2MASS and SSS astrometry, assuming positional uncertainties of 0$\farcs$3 \citep{ham01a,cut03}.}
\tablenotetext{e}{Original discovery reference for each source.}
\tablerefs{(1) \citet{mewide3}; (2) This paper; (3) \citet{sch1013}; (4) \citet{me1626}}
\end{deluxetable}

\begin{deluxetable}{lccccl}
\tabletypesize{\footnotesize}
\tablecaption{Observation Log. \label{tab_log}}
\tablewidth{0pt}
\tablehead{
\colhead{Source} &
\colhead{Instrument} &
\colhead{UT Date} &
\colhead{t$_{int}$ (s)} &
\colhead{$\sec{z}$} &
\colhead{Calibrator\tablenotemark{a}} \\
}
\startdata
2MASS 0142+0523 & GMOS & 2004 Oct 8 & 3600 & 1.03 & \nodata  \\
2MASS 0851-0005 & GMOS & 2004 Nov 20 & 3600 & 1.55 & \nodata  \\
2MASS 0934+0536 & GMOS & 2004 Nov 20 & 3600 & 1.11 & \nodata  \\
SSSPM 1013-1356\tablenotemark{b} & GMOS & 2004 Nov 21 & 1800 & 1.27 & \nodata  \\
 & LDSS-3 & 2006 May 8 & 750 & 1.07 & HD 89013  \\
2MASS 1227-0447 & LDSS-3 & 2006 May 7 & 600 & 1.10 & G 104-335 \\
2MASS 1626+3925 & GMOS & 2004 Aug 17 & 3600 & 1.36 & \nodata \\
2MASS 1640+1231 & LDSS-3 & 2006 May 8 & 1200 & 1.35 & HD 154086 \\
\enddata
\tablenotetext{a}{G dwarf star observed for telluric calibration.}
\tablenotetext{b}{Observed with both GMOS and LDSS-3.}
\end{deluxetable}

\begin{deluxetable}{lccccccccc}
\tabletypesize{\scriptsize}
\tablecaption{Atomic Line Equivalent Widths ({\AA}).\label{tab_ews}}
\tablewidth{0pt}
\tablehead{
\colhead{Source} &
\colhead{H$\alpha$} &
\colhead{{\ion{Ca}{1}}} &
\colhead{{\ion{Ti}{1}}} &
\colhead{{\ion{Rb}{1}}} &
\colhead{{\ion{Rb}{1}}} &
\colhead{{\ion{Na}{1}}} &
\colhead{{\ion{Ti}{1}}} &
\colhead{{\ion{Cs}{1}}} &
\colhead{{\ion{Ca}{2}}} \\
\colhead{} &
\colhead{(6563 {\AA})} &
\colhead{(6573 {\AA})} &
\colhead{(7209/7213 {\AA})} &
\colhead{(7800 {\AA})} &
\colhead{(7948 {\AA})} &
\colhead{(8183/8195 {\AA})} &
\colhead{(8436 {\AA})} &
\colhead{(8521 {\AA})} &
\colhead{(8542 {\AA})}  \\
}
\startdata
2MASS 0934+0536 & -1.5$\pm$0.5 & $<$1.1 & 4.6$\pm$1.0 & 0.9$\pm$0.3 & 0.9$\pm$0.4 & 6.9$\pm$0.5 & 1.3$\pm$0.3  & $<$0.2 & $<$0.2  \\
2MASS 0851-0005 & -1.0$\pm$0.5 &  $<$1.9 & $<$2.0 & $<$1.5 & 1.2$\pm$0.3 & 7.6$\pm$0.5 & 1.5$\pm$0.4  & $<$1.0 & $<$0.5 \\
2MASS 1640+1231 & $>-$0.4 & 2.4$\pm$0.2 & 4.0$\pm$0.4 & 2.0$\pm$0.2  & 2.2$\pm$0.3 & 10.0$\pm$0.5 & 1.8$\pm$0.2 & $<$0.5 & 1.1$\pm$0.2 \\
2MASS 0142+0523 & $>-$0.2 & 3.9$\pm$0.4 & 3.4$\pm$0.3 & 1.1$\pm$0.2 & 1.1$\pm$0.2  & 9.5$\pm$0.4 & 2.1$\pm$0.3 & $<$0.3 & 2.6$\pm$0.3  \\
SSSPM 1013-1356 & $>-$0.2 & 3.5$\pm$0.4 & 3.8$\pm$0.3 & 1.5$\pm$0.2 & 1.3$\pm$0.2 & 10.0$\pm$0.4 & 2.4$\pm$0.4 & $<$0.3 & 2.3$\pm$0.2  \\
2MASS 1626+3925 & $>-$0.4 & 4.0$\pm$1.0 & $<$0.3 & 4.6$\pm$0.2 & 3.8$\pm$0.4  & 9.0$\pm$1.0 & 2.5$\pm$0.5 & 1.8$\pm$0.3 & $<$0.3  \\
2MASS 1227-0447 & $>-$0.3 & 2.4$\pm$0.2 & 2.9$\pm$0.5 & 0.9$\pm$0.2  & 0.9$\pm$0.2  & 6.7$\pm$0.4 & 2.0$\pm$0.4 & $<$0.2 & 3.6$\pm$0.3  \\
\enddata
\end{deluxetable}

\begin{deluxetable}{lcccccccccl}
\tabletypesize{\footnotesize}
\tablecaption{Spectral Ratios and Classifications.\label{tab_class}}
\tablewidth{0pt}
\tablehead{
\colhead{Object} &
\colhead{TiO5} &
\colhead{CaH1} &
\colhead{CaH2} &
\colhead{CaH3} &
\colhead{VO1} &
\colhead{TiO6} &
\colhead{VO2} &
\colhead{TiO7} &
\colhead{Color-M} &
\colhead{SpT} \\
}
\startdata
2MASS 0934+0536 & 0.211 & 0.357 & 0.203 & 0.505 & 0.836 & 0.439 & 0.482 & 0.667 & 6.406 & M7  \\
 & [M5.7] &  & [M6.9] & [M6.7] & M6.7 & M7.0 & M7.3 & M6.4 & M7.6 & \\
2MASS 0851-0005 & 0.194 & 0.365 & 0.227 & 0.509 & 0.830 & 0.412 & 0.374 & 0.560 & 10.163 &  M8 pec \\
 & [M5.9] &  &  [M6.4] & [M6.6] & M6.9 & M7.3 & M8.5 & M7.5 & M9.1 &  \\
2MASS 1640+1231 & 0.078 & 0.172 & 0.113 & 0.244 & 0.951 & 0.499 & 0.621 & 0.703 & 6.895 &  d/sdM9 \\
 &  &  &  sdM8.5 & sdM9.9 &  & & &  &  [sdL3.5] & \\
2MASS 0142+0523 & 0.218 & 0.203 & 0.158 & 0.290 & 0.948 & 0.832 & 0.797 & 0.914 & 3.079 & sdM8.5  \\
 &  &  &  sdM7.7 & sdM9.1 &  &  &  &  &  [sdM7.6]  & \\
SSSPM 1013-1356 & 0.248 & 0.131 & 0.114 & 0.204 & 0.983 & 0.982 & 0.855 & 0.968 & 2.866 &  sdM9.5 \\
 &  &  &  sdM8.5 & sdL0.5 &  &  &  &  &  [sdM7.1]  &  \\
(LDSS-3) & 0.248 & 0.121 & 0.122 & 0.198 & 0.977 & 0.956 & 0.868 & 0.957 & 2.931  &  sdM9.5 \\
 &  &  &  sdM8.3 & sdL0.6 &  &  &  &  &  [sdM7.3]  &  \\
2MASS 1626+3925 & 0.260 & 0.192 & 0.098 & 0.131 & 1.504 & 0.691 & 0.602 & 0.982 & 5.603 &  sdL4\tablenotemark{a}  \\
 &  &  &  [sdM8.8] & [sdL1.7] &  &  &  &  &  [sdL2.0]  &  \\
2MASS 1227-0447 & 0.825 & 0.114 & 0.177 & 0.265 & 0.923 & 1.162 & 0.966 & 1.027 & 2.379 &  esdM7.5 \\
 &  &  &  esdM7.3 & esdM7.9 &  &  &  &  &  [esdL5.0]   &  \\
\enddata
\tablecomments{Index spectral types based on the relations defined in \citet{giz97}
and \citet{lep03}.  Final spectral types (SpT) are the average of the index types, with the
exception of those listed in brackets.}
\tablenotetext{a}{Based on the similarity of the spectrum of this source 
to the L4 spectral standard; see
$\S$~4.5.}
\end{deluxetable}

\begin{deluxetable}{lcccccc}
\tabletypesize{\footnotesize}
\tablecaption{Distance Estimates and Kinematics. \label{tab_kinematics}}
\tablewidth{0pt}
\tablehead{
\colhead{Source} &
\colhead{$d_{est}$} &
\colhead{$V_{tan}$} &
\colhead{$V_{rad}$} &
\colhead{$U$} &
\colhead{$V$} &
\colhead{$W$} \\
\colhead{} &
\colhead{(pc)} &
\colhead{({\kms})} &
\colhead{({\kms})} &
\multicolumn{3}{c}{({\kms})} \\
}
\startdata
2MASS 0142+0523 & 65 & 194 & 63$\pm$11 & -114 & -137 & -58   \\
2MASS 0851-0005 & 110 & 73 & 72$\pm$24 & -23 & -83 & -16  \\
2MASS 0934+0536 & 105 & {\nodata} & 156$\pm$28 & \nodata & \nodata & \nodata  \\
SSSPM 1013-1356 & 30 & 146 & 50$\pm$7\tablenotemark{a} & 87 & -110 & -48 \\
2MASS 1227-0447 & 65 & 148 & 91$\pm$9 & 103 & 83 & 135 \\
2MASS 1626+3925 & 20 & 120 & -213$\pm$70\tablenotemark{b} & -36 & -132 & -213 \\
2MASS 1640+1231 & 75 & 291 & -46$\pm$8 & -7 & -250 & 143 \\
\enddata
\tablenotetext{a}{Based on LDSS-3 data.}
\tablenotetext{b}{This measurement appears to be affected by systematic errors in the wavelength calibration.
The $V_{rad} = -260{\pm}35$~{\kms} measured by \citet{me1626} was used to compute $UVW$ velocities.}
\end{deluxetable}

\begin{deluxetable}{llcccc}
\tabletypesize{\footnotesize}
\tablecaption{Spectral Model Fit Parameters.\tablenotemark{a} \label{tab_modelfits}}
\tablewidth{0pt}
\tablehead{
\multicolumn{2}{c}{} &
\multicolumn{2}{c}{COND} &
\multicolumn{2}{c}{NextGen} \\
\cline{3-4} \cline{5-6}
\colhead{Source} &
\colhead{SpT} &
\colhead{{\teff}} &
\colhead{[M/H]} &
\colhead{{\teff}} &
\colhead{[M/H]} \\
\colhead{} &
\colhead{} &
\colhead{(K)} &
\colhead{(dex)} &
\colhead{(K)} &
\colhead{(dex)} \\
}
\startdata
2MASS 0934+0536 & M7 & 2780$\pm$90 & -0.2$\pm$0.3 & 2890$\pm$90 & -0.2$\pm$0.2  \\
2MASS 0851-0005 & M8 pec & 2720$\pm$90 & -0.2$\pm$0.2 & 2830$\pm$100 & -0.2$\pm$0.2  \\
2MASS 1640+1231 & d/sdM9 & 2720$\pm$80 & -0.2$\pm$0.3 & 2860$\pm$70 & -0.5$\pm$0.4 \\
2MASS 0142+0523 & sdM8.5 & 2910$\pm$60 & -0.5$\pm$0.4 & 3000$\pm$70 & -0.6$\pm$0.4  \\
SSSPM 1013-1356 & sdM9.5 & 2850$\pm$80 & -1.0$\pm$0.4 & 2930$\pm$90 & -1.1$\pm$0.3  \\
2MASS 1626+3925\tablenotemark{b} & sdL4 & 2410$\pm$90 & -1.3$\pm$0.3 & 2380$\pm$140 & -1.8$\pm$0.2 \\
2MASS 1227-0447 & esdM7.5 & 3080$\pm$70 & -1.3$\pm$0.4 & 3130$\pm$70 & -1.4$\pm$0.3 \\
\enddata
\tablenotetext{a}{Uncertainties are based on the scatter of values from 
the eight best spectral fits weighted by their ${\chi}^2$ residuals.
Systematic uncertainties are not included.}
\tablenotetext{b}{Parameters are highly uncertain for this source due to the
poor quality of the spectral model fits (see Figure~\ref{fig_modelfits2}).}
\end{deluxetable}

\clearpage
\thispagestyle{empty}
\setlength{\hoffset}{-15mm}
\begin{deluxetable}{llccccccccl}
\tabletypesize{\scriptsize}
\tablecaption{Spectroscopically Confirmed Ultracool Subdwarfs.\label{tab_compendium}}
\tablewidth{0pt}
\tablehead{
\colhead{Source} & 
\colhead{SpT} & 
\colhead{$R_{ESO}$\tablenotemark{a}} & 
\colhead{$R_{ESO}-I_N$\tablenotemark{a}} & 
\colhead{$I_N-J$\tablenotemark{a,b}} & 
\colhead{$J$\tablenotemark{b}} & 
\colhead{$J-K$\tablenotemark{b}} & 
\colhead{$\mu$} &
\colhead{$V_r$} &
\colhead{$d_{est}$} &
\colhead{Ref.} \\
\colhead{} & 
\colhead{} & 
\colhead{} & 
\colhead{} & 
\colhead{} & 
\colhead{} & 
\colhead{} & 
\colhead{($\arcsec$ yr$^{-1}$)} & 
\colhead{({\kms})} & 
\colhead{(pc)} & 
\colhead{} \\ 
}
\startdata
LSR 1610-0040 & d/sdM7: &  17.51 & 2.7 & 1.8 & 12.99$\pm$0.02 & 0.97$\pm$0.03 & 1$\farcs$46 & -95   &  16 & 1,2,3 \\
SSSPM 1444-2019 & d/sdM9 & 18.57 & 3.6 & 2.4 &  12.55$\pm$0.03 & 0.61$\pm$0.15 & 3$\farcs$51 & -156 & 20 & 4 \\
2MASS 1640+1231 & d/sdM9 & 20.63 & \nodata & \nodata &  15.95$\pm$0.08 & $<$0.4 & 0$\farcs$82 & -46 & 75 & 5 \\
2MASS 0937+2931 & d/sdT6 & \nodata & \nodata & \nodata &  14.64$\pm$0.04 & -0.62$\pm$0.13 & 1$\farcs$62 & \nodata & 6.1\tablenotemark{c} & 6,7 \\
LHS 377 & sdM7 & 17.34 & 2.7 & 1.5 & 13.19$\pm$0.03 & 0.71$\pm$0.04 & 1$\farcs$25 & 180 & 35\tablenotemark{c} & 8,9,10,11  \\
SSSPM 1930-4311 & sdM7 & 18.42 & 2.1 & 1.6 & 14.79$\pm$0.03 & 0.70$\pm$0.07 & 0$\farcs$87 & -262 & 73  & 12 \\
LSR 2036+5059 & sdM7.5 & 17.46 & 2.2 & 1.6 & 13.61$\pm$0.03 & 0.68$\pm$0.04 &  1$\farcs$05 & -140  & 18 & 13 \\
LSR 1425+7102 & sdM8 & 18.61 & 2.4 & 1.4 & 14.78$\pm$0.04 & 0.45$\pm$0.10 &  0$\farcs$64 & -65 &  65 & 14 \\
2MASS 0142+0523  & sdM8.5 & \nodata & \nodata & \nodata &  15.91$\pm$0.08 & 0.31$\pm$0.24 & 0$\farcs$63 & 63 & 65 & 5,15 \\
SSSPM 1013-1356 & sdM9.5 & 18.69 & 2.6 & 1.8 & 14.59$\pm$0.04 & 0.24$\pm$0.09 & 1$\farcs$03 & 50 & 30 & 5,12 \\
SDSS 1256-0224 & sdL4: & \nodata & \nodata  & 2.3 & 16.10$\pm$0.11 & $<$0.7 & 0$\farcs$62 & -95 & 120 & 16 \\
2MASS 1626+3925 & sdL4 & 19.84 & 3.2 & 2.3 & 14.44$\pm$0.03 &  -0.03$\pm$0.08 & 1$\farcs$27 & -260 &  20 & 5,17 \\
2MASS 0532+8246 & sdL7 & \nodata & \nodata & \nodata & 15.18$\pm$0.06 & 0.26$\pm$0.16 &  2$\farcs$60 & -172 & 20 & 3,18 \\
APMPM 0559-2907 & esdM7 & 18.12 & 1.8 & 1.4 & 14.89$\pm$0.04 & 0.43$\pm$0.08 & 0$\farcs$38 & 180 & 70 & 19,20 \\
2MASS 1227-0447 & esdM7.5 & 18.48 & 1.9 & 1.1 & 15.49$\pm$0.05 & 0.61$\pm$0.13 & 0$\farcs$48 & 91 & 65 & 5 \\
LEHPM 2-59 & esdM8 & 18.82 & 1.6 & 1.7 & 15.52$\pm$0.05 & 0.76$\pm$0.12 & 0$\farcs$75 & 79  & 66 &  21,22 \\
\enddata
\tablecomments{Includes sources spectral types d/sdM7, sdM7 and esdM7 
and later \citep{mecs13} based on reported optical spectroscopy.}
\tablenotetext{a}{Photographic plate photometry from the SuperCOSMOS Sky Survey \citep{ham01a,ham01b,ham01c}.}
\tablenotetext{b}{Near-infrared photometry from 2MASS \citep{skr06}.}
\tablenotetext{c}{Parallax distance measurement \citep{mon92,vrb04}.}
\tablerefs{(1) \citet{lep1610}; (2) \citet{cus06}; (3) \citet{rei06}; 
(4) \citet{sch1444}; (5) This paper; (6) \citet{me02a}; (7) \citet{metgrav};
(8) \citet{luy79a}; (9) Boeshaar \& Liebert (1991), private communication, as
reported in \citet{mon92}; (10) \citet{mon92}; (11) \citet{giz97};
(12) \citet{sch1013}; (13) \citet{lep03}; (14) \citet{lep1425}; (15) \citet{mewide3}; 
(16) \citet{siv06}; 
(17) \citet{me1626}; (18) \citet{me0532}; (19) \citet{sch99}; (20) \citet{lep0822};
(21) \citet{por04}; (22) \citet{melehpm2-59}.}
\end{deluxetable}

\clearpage
\setlength{\hoffset}{0mm}

\clearpage

\begin{figure}
\centering
\caption{POSS-II and SERC $R$-band (left) and 2MASS $J$- (middle) and $K_s$-band (right)
images of 2MASS 0142+0523, 2MASS 0851-0005, 2MASS 0934+0536, 2MASS 1227-0447
and 2MASS 1640+1231.
Images are scaled to the same spatial resolution, 
5$\arcmin$ on a side, with north up and east to the left.
Inset boxes 20$\arcsec$$\times$20$\arcsec$ in size indicate 
the 2MASS position of the source, and are expanded
in the lower left corner of each image. [THESE IMAGES ARE SEPARATELY INCLUDED IN PDF FORMAT IN THE SOURCE FILES]
\label{fig_finders}}
\end{figure}

\clearpage
\centering
\centerline{Fig.\ 1. --- Continued. [THESE IMAGES ARE SEPARATELY INCLUDED IN PDF FORMAT IN THE SOURCE FILES]}

\clearpage

\begin{figure}
\centering
\epsscale{0.8}
\plotone{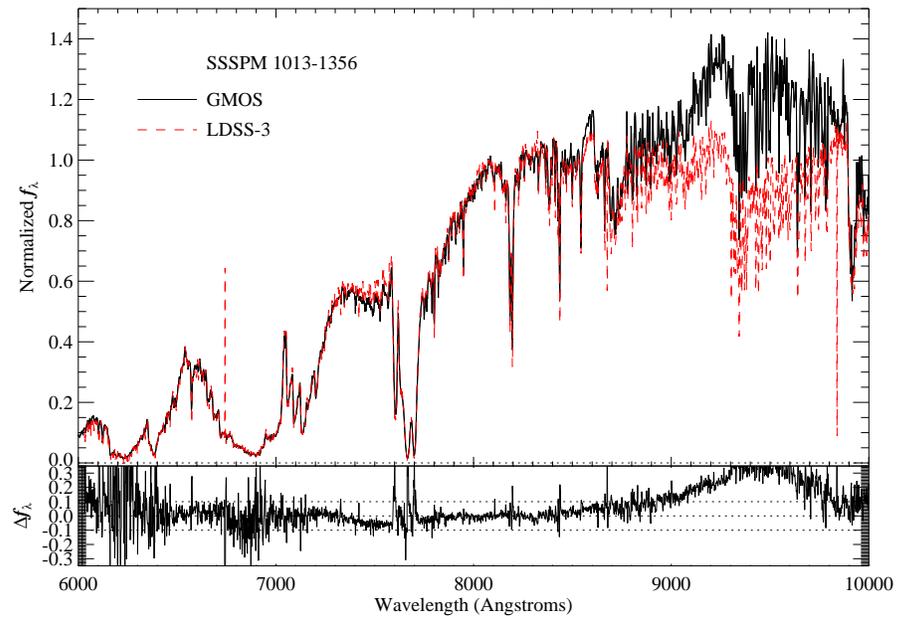}
\caption{Top panel shows a 
comparison of optical spectra for SSSPM~1013-1356 obtained
with GMOS (solid black line) and LDSS-3 (red dashed line,
without telluric correction).
Both spectra are normalized 8075 {\AA}.
The bottom panel shows the fractional deviation between the spectra,
with the 0\% and $\pm$10\% deviation indicated by dotted lines.
\label{fig_gmosvsldss3}}
\end{figure}

\begin{figure}
\centering
\epsscale{0.8}
\plotone{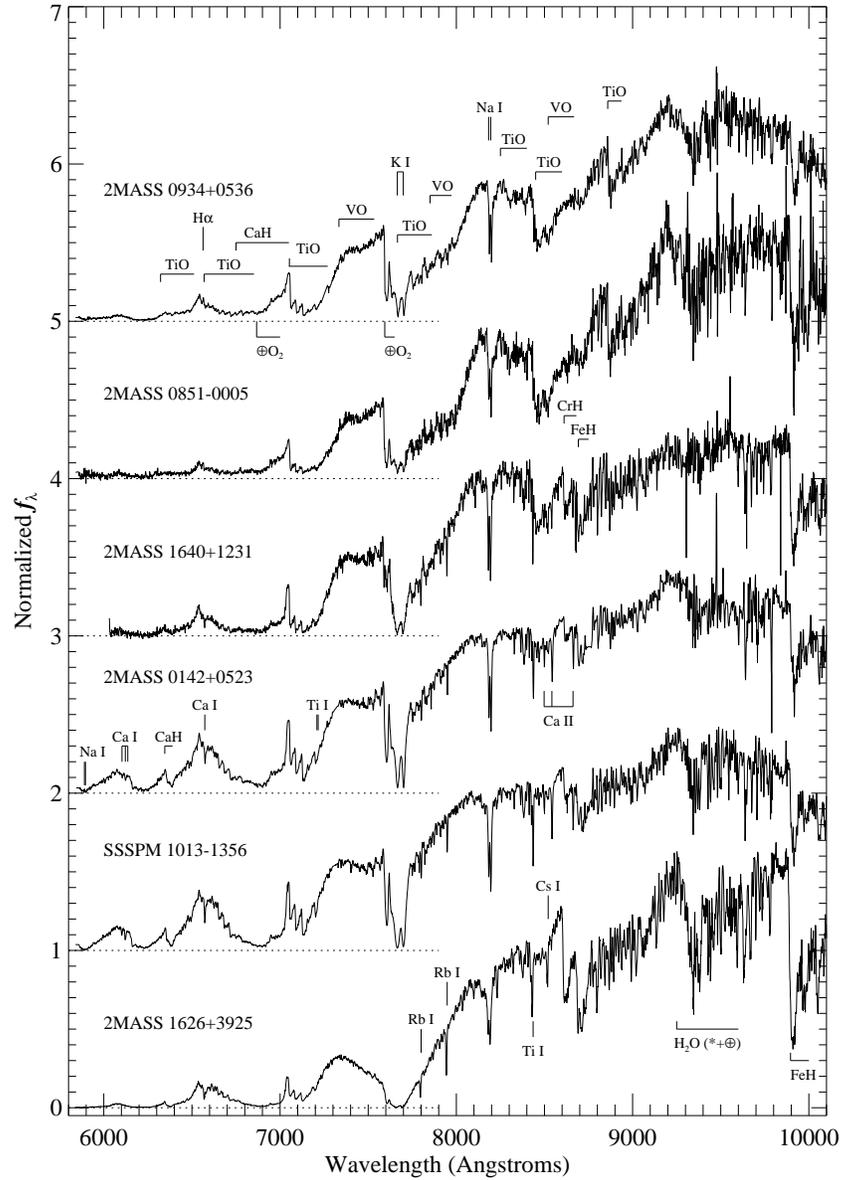}
\caption{Optical spectra of (from top to bottom) 
2MASS 0934+0536, 2MASS 0851-0005, 2MASS 1640+1231, 2MASS 0142+0523,
SSSPM 1013-1356 and 2MASS 1626+3925.  
Data are normalized 8075 {\AA}
and offset by constants (dotted lines).
All spectra are from GMOS observations
with the exception of 2MASS 1640+1231, for which
LDSS-3 data are shown.
Identified molecular features are from
\citet{kir91} and \citet{kir99}; atomic features are from \citet{kur95}. 
Features arising from telluric H$_2$O and O$_2$ absorption in the GMOS data
are indicated by $\oplus$ symbols (LDSS-3 data for 2MASS 1640+1231 have been
corrected for telluric absorption).
\label{fig_spectra}}
\end{figure}

\begin{figure}
\centering
\epsscale{0.8}
\plotone{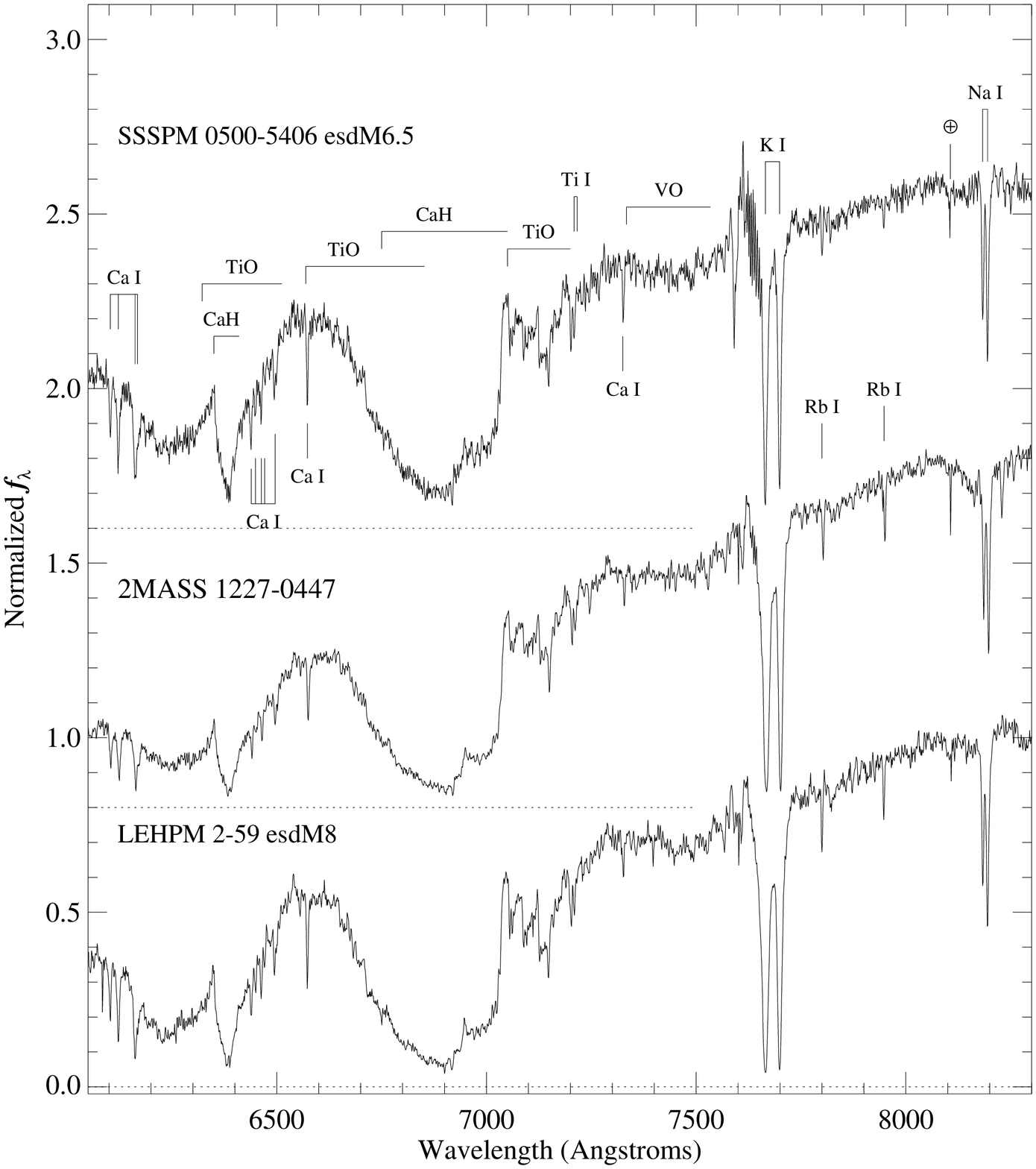}
\caption{LDSS-3 optical spectrum of 2MASS 1227-0447 (middle) compared
to those of the esdM6.5 SSSPM 0500-5406 (top) and the esdM8 LEHPM 2-59 (bottom;
data from \citet{melehpm2-59}).
All data are telluric corrected, normalized at 8100 {\AA}
and offset by constants (dotted lines).
Identified molecular and atomic features are from
\citet{kir91,kir99}; and \citet{kur95}. 
The noisy section around 7600 {\AA} in the spectrum of SSSPM 0500-5406
arises from poor correction of telluric O$_2$ absorption.
\label{fig_esdm}}
\end{figure}

\begin{figure}
\centering
\epsscale{0.8}
\plotone{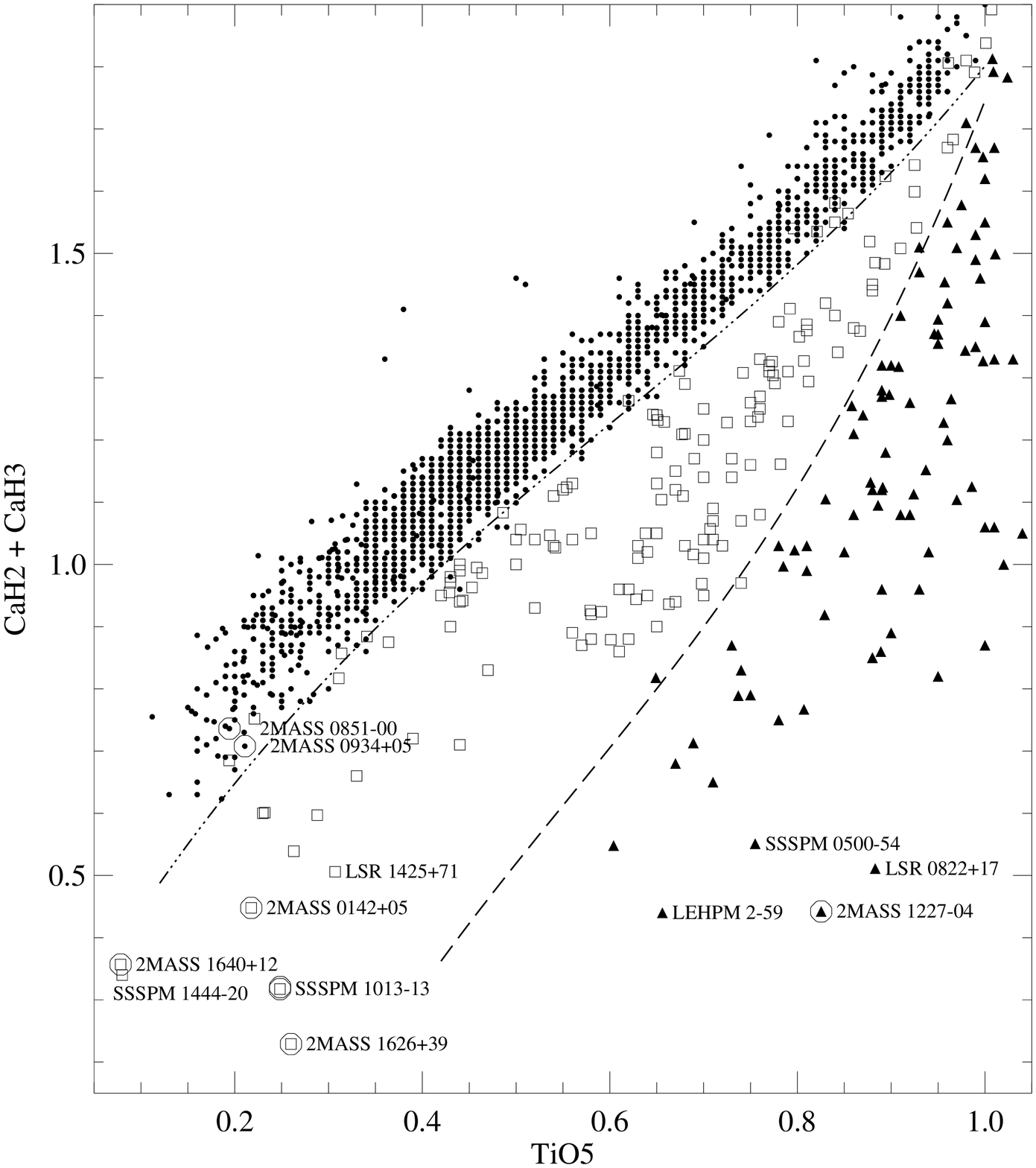}
\caption{Spectral indices CaH2+CaH3 versus TiO5 for dwarfs (points), 
subdwarfs (open squares) and extreme subdwarfs (filled triangles)
from \citet{haw96,giz97,giz97b,rei02,lep03,lep1425,lep0822,sch1013,sch1444}; and \citet{melehpm2-59}.
Data from this paper
are encircled and labeled, while
values for the sdM8 LSR 1425+7102, the d/sdM9 SSSPM 1444-2019,
the esdM6.5s LSR 0822+1700 and SSSPM 0500-5406, and
the esdM8 LEHPM 2-59 are also labeled.
Dashed and dot-dashed lines delineate boundaries between dwarfs, subdwarfs and 
extreme subdwarfs as defined by \citet{melehpm2-59}.
\label{fig_cah23tio}}
\end{figure}

\begin{figure}
\centering
\epsscale{0.8}
\plotone{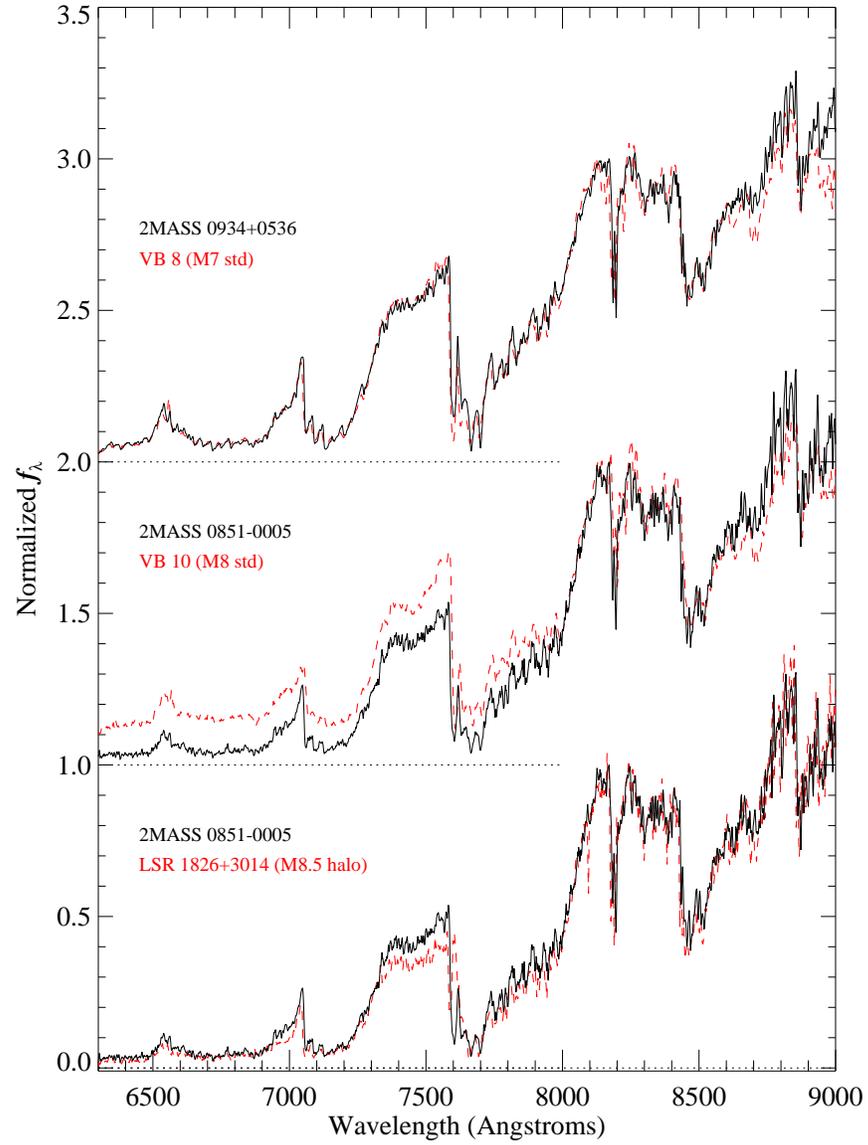}
\caption{Comparison of the 6300--9000 {\AA} optical spectra
of (black lines) 2MASS 0934+0536 (top) and 2MASS 0851-0005 (middle and bottom)
to spectral standards (red dashed lines) VB 8 (M7; top), VB 10 (M8; middle) and
the high proper motion M8.5 LSR 1826+3014 (bottom).
Data for VB~8 and VB~10 are from \citet{cru03}; data for LSR 1826+3014
are from \citet{lep1826}.
All spectra are normalized at 8100 {\AA} and offset by
a constant (dotted lines).  Note the excellent match between
2MASS 0934+0536 and VB~8, but relatively poor match between 
2MASS 0851-0005 and VB~10 shortward of 8000 {\AA}.
There is better overall agreement between 2MASS 0851-0005 and LSR 1826+3014.
\label{fig_mdwcomp}}
\end{figure}

\begin{figure}
\centering
\epsscale{0.8}
\plotone{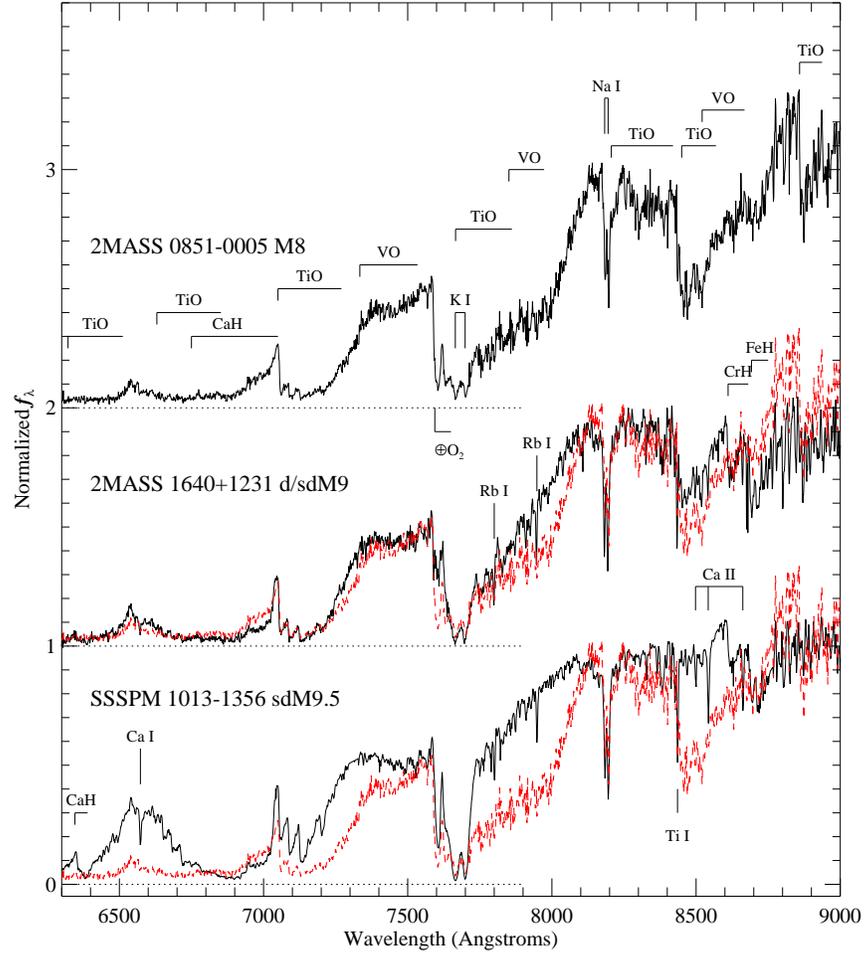}
\caption{Examination of metallicity effects in the 6300--9000 {\AA} spectral region.
Shown in black from top to bottom are data for 2MASS 0851-0005, 2MASS 1640+1231 
and SSSPM 1013-1356 (bottom); the spectrum of 2MASS 0851-0005 is also overlain on those
of the other two sources (red dashed lines).  All spectra are normalized at 8100 {\AA},
corrected for radial motion and offset
by constants (dotted lines).  The spectrum of 2MASS 1640+1231 has 
been corrected for telluric absorption.
Key spectral features are indicated.  Note in particular how metal oxide bands at 
7200, 7900 and 8500 {\AA} weaken down the sequence, revealing numerous atomic lines of
\ion{Ca}{1}, \ion{Ca}{2}, \ion{Rb}{1} and \ion{Ti}{1}.  Note also the similarity in {\ki} and
{\nai} line strengths between the three sources.
\label{fig_zcomp}}
\end{figure}

\begin{figure}
\centering
\epsscale{0.8}
\plotone{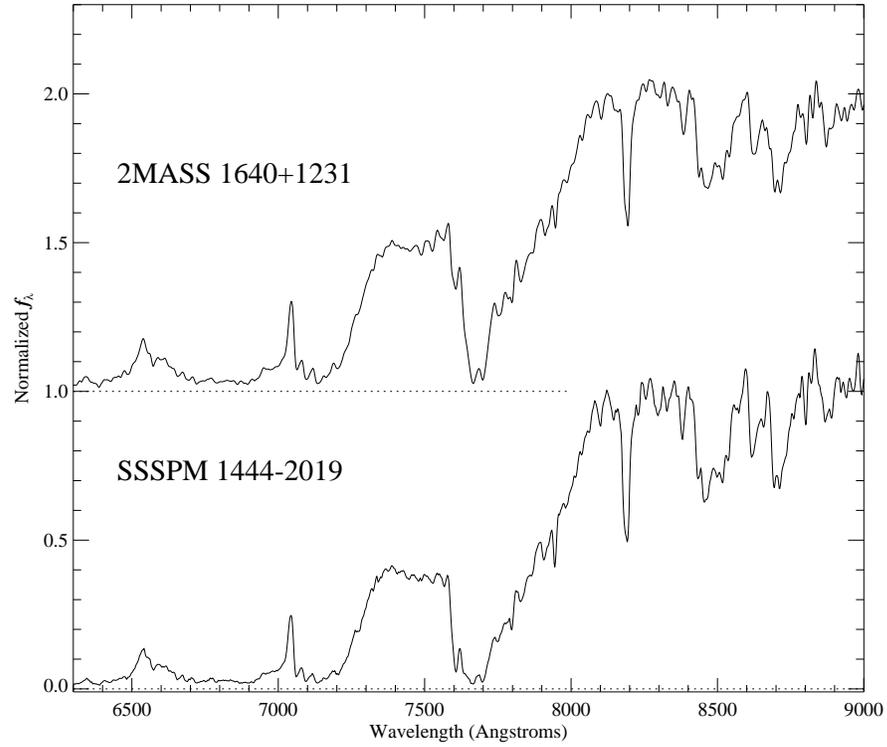}
\caption{Comparison of the 6300--9000 {\AA} optical spectra of 2MASS 1640+1231 (top)
and SSSPM 1444-2019 (bottom, data from \citet{sch1444}).
Both spectra are normalized at 8100 {\AA}, corrected for radial motion 
and vertically offset (dotted lines).  Data for 2MASS~1640+1231 have
been deconvolved to the resolution of the specrtum of SSSPM 1444-2019
({\ldl} $\sim$ 550) using a Gaussian kernel.  The latter spectrum
has not been corrected for telluric 
absorption.
\label{fig_1640comp}}
\end{figure}

\begin{figure}
\centering
\epsscale{0.8}
\plotone{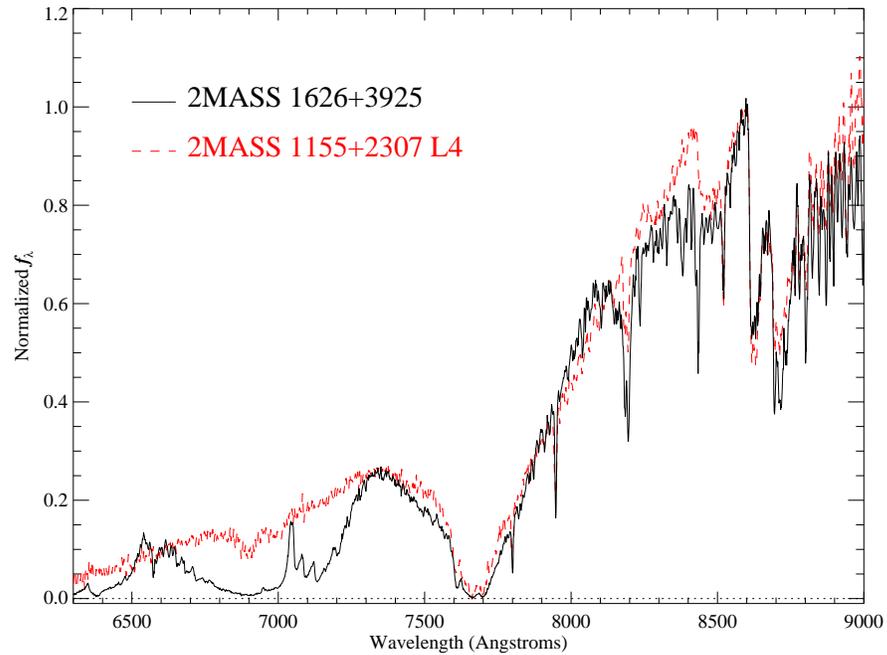}
\caption{Comparison of the 6300--9000 {\AA} optical spectra of 2MASS 1626+3925 (black line)
and the L4 field dwarf 2MASS 1155+2307 (red dashed line; data from \citet{kir99}).
Both spectra are normalized at 8600 {\AA}, and the spectrum of
2MASS 1626+3925 has been corrected for radial motion.  Note the fairly good agreement between
the spectra over  7300 to 9000 {\AA}, with the exception of stronger
TiO absorption over 8300--8450 {\AA} and stronger \ion{Na}{1} lines in 
the spectrum of 2MASS 1626+3925.
\label{fig_2m1626comp}}
\end{figure}

\begin{figure}
\centering
\epsscale{1.0}
\plotone{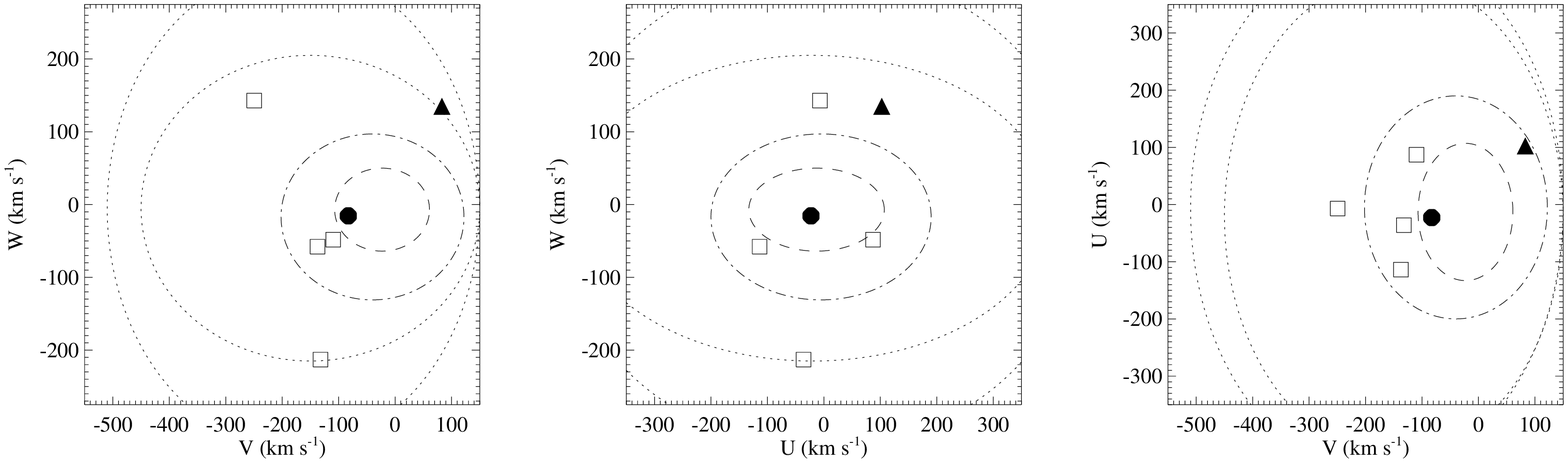}
\caption{$UVW$ kinematics in the local standard of rest frame for sources observed
in this study.  Kinematics for the M dwarf 2MASS 0851-0005 and esdM 2MASS 1227-0447 
are indicated by the solid disks and triangles, respectively; the subdwarfs
are indicated by open squares.  The 3$\sigma$ velocity dispersion 
spheres of local disk M dwarfs \citep[dashed lines]{haw96},
thick disk stars \citep[dot-dashed lines]{str87} and halo stars 
\citep[dotted lines, spherical and flat components]{som90} are indicated.
\label{fig_uvw}}
\end{figure}

\begin{figure}
\centering
\includegraphics[width=0.48\textwidth]{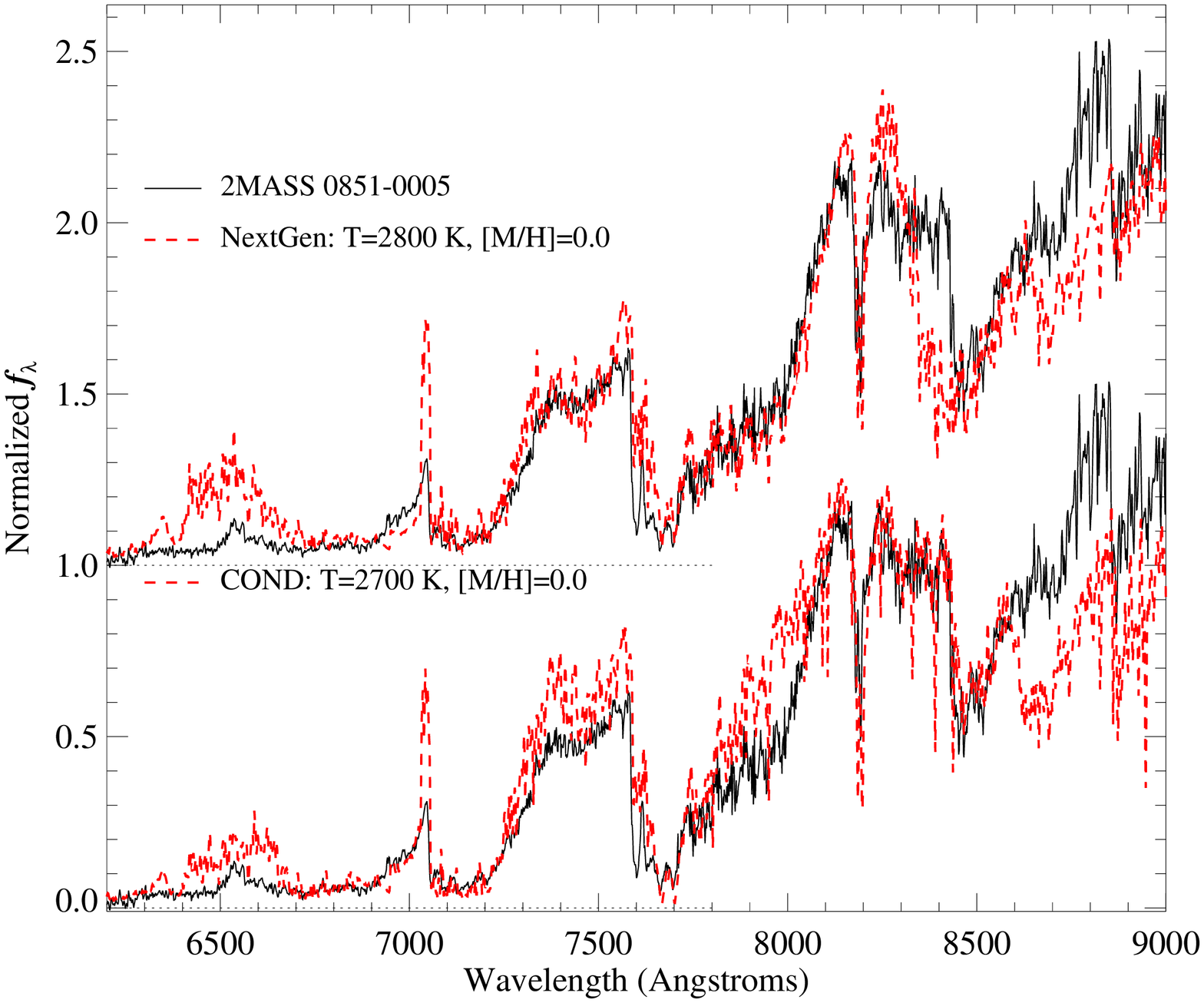}
\includegraphics[width=0.48\textwidth]{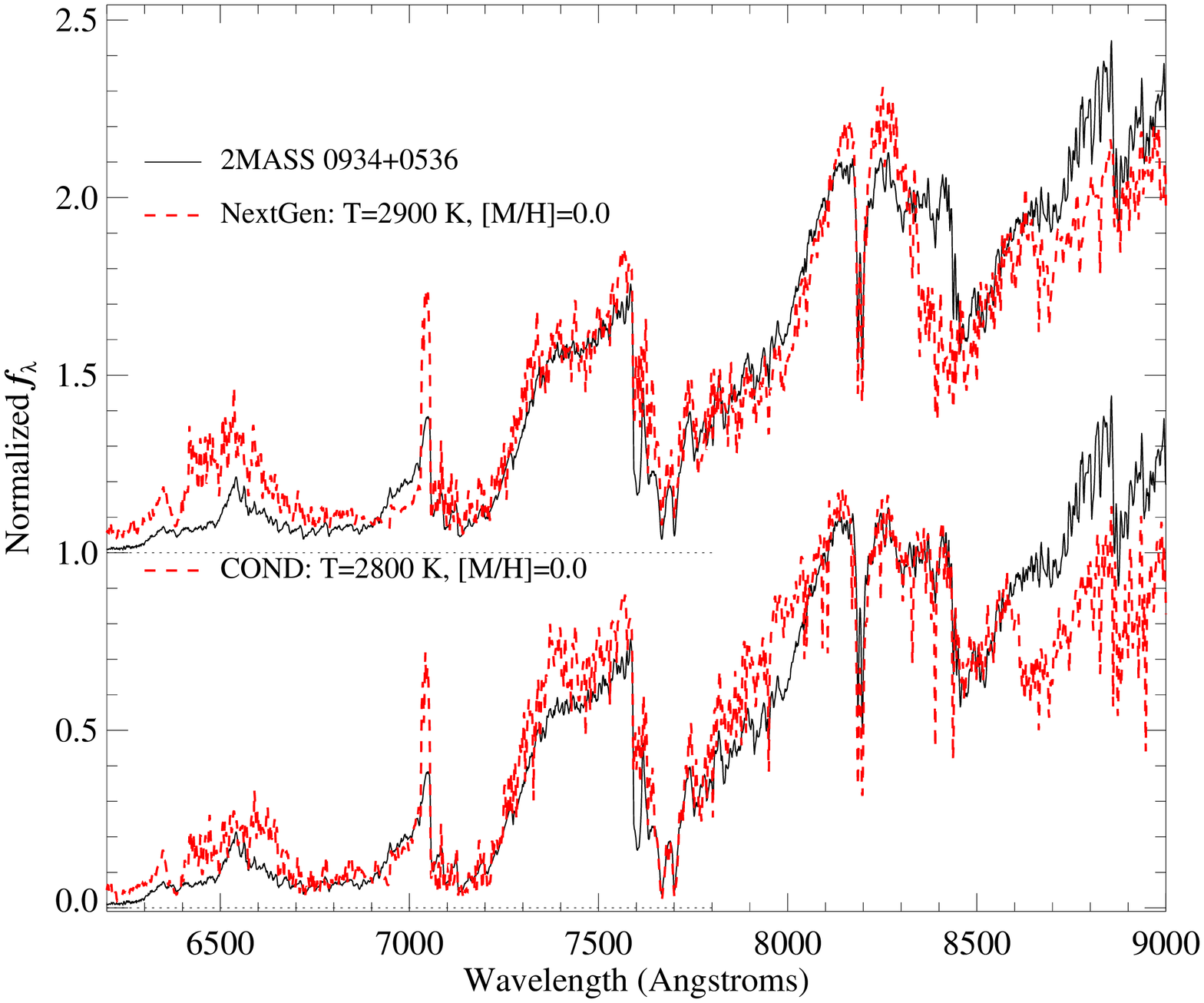}
\includegraphics[width=0.48\textwidth]{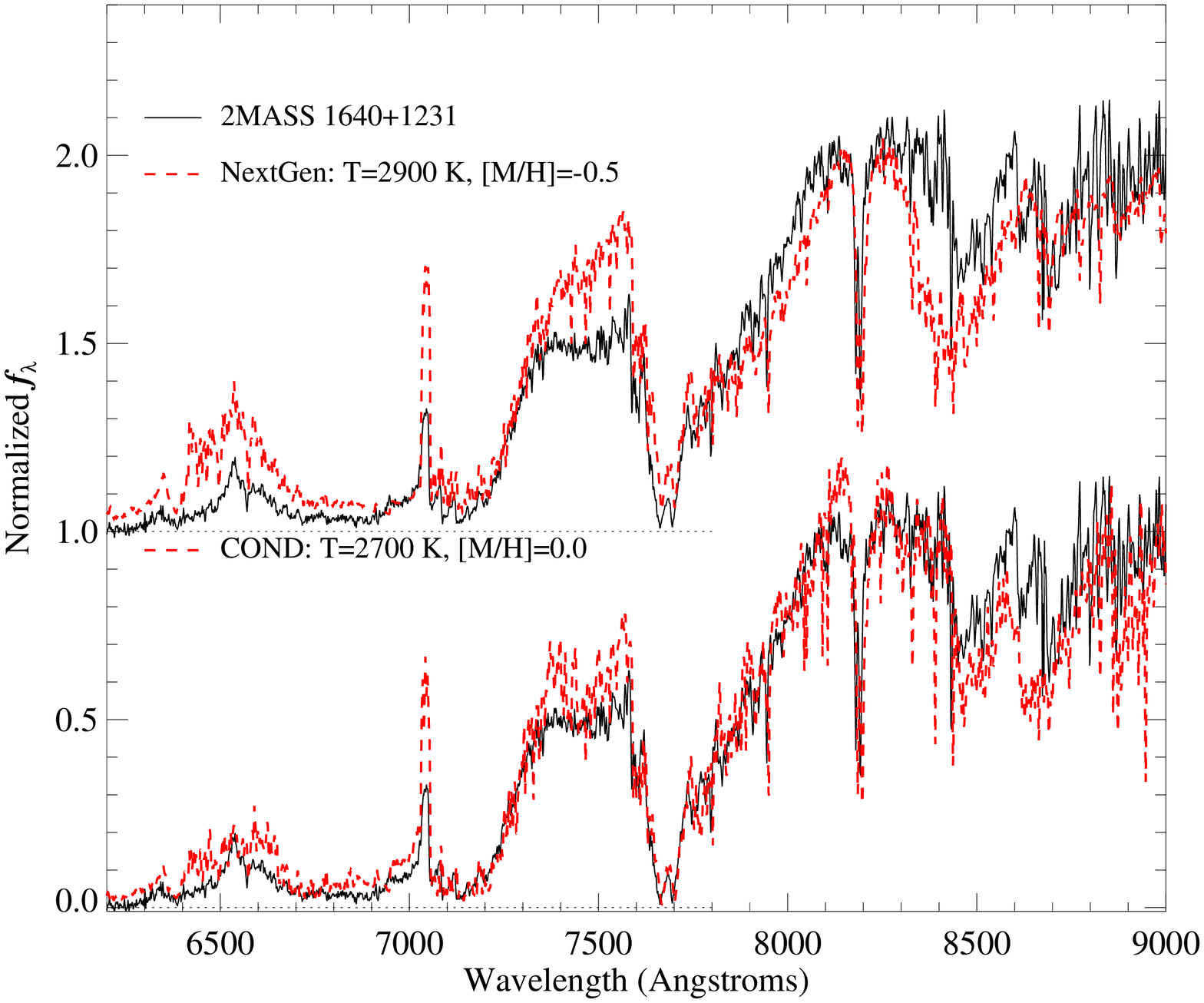}
\includegraphics[width=0.48\textwidth]{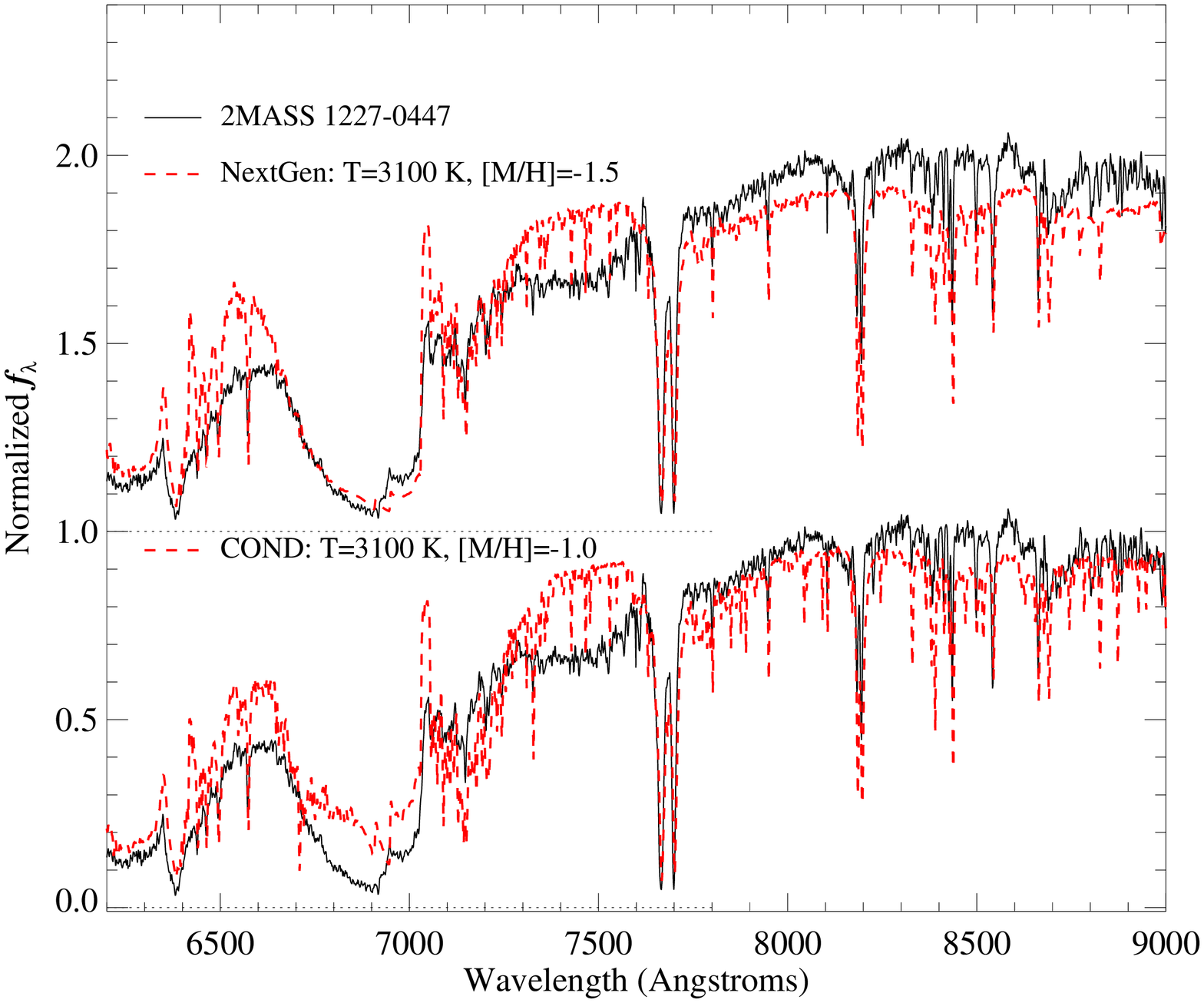}
\caption{Best fit NextGen (top spectrum) and COND (bottom spectrum) model
fits for the M dwarfs 2MASS 0851-0005 and 2MASS 0934+0536, the mild subdwarf
2MASS 1640+1231, and the extreme subdwarf 2MASS 1227-0447.  
Each panel displays the normalized
spectrum of the source (black solid lines) overlaid with the best fitting model
(red dashed lines).  The models assume $\log{g} = 5.5$ (cgs).
\label{fig_modelfits}}
\end{figure}

\begin{figure}
\clearpage
\centering
\includegraphics[width=0.48\textwidth]{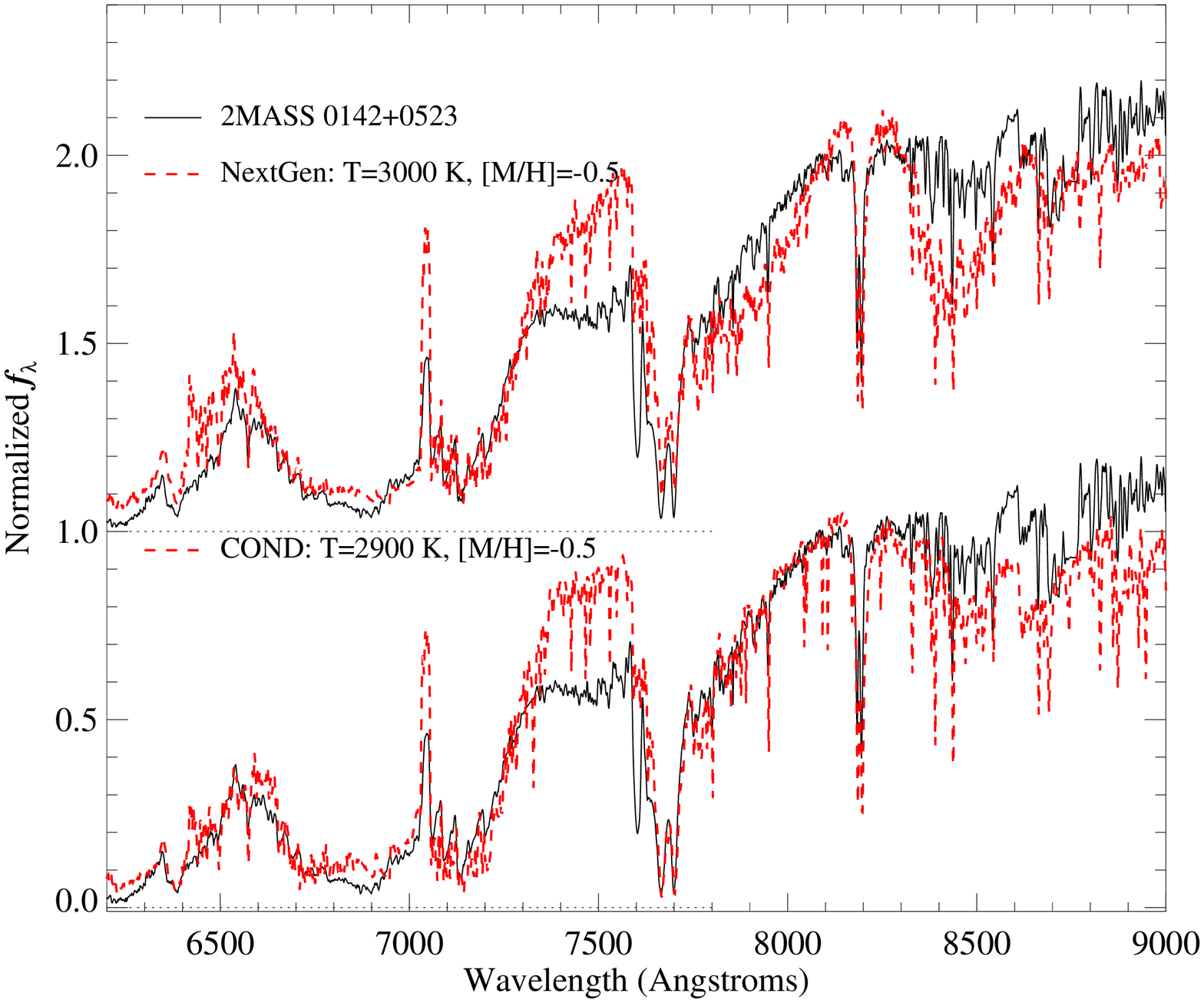}
\includegraphics[width=0.48\textwidth]{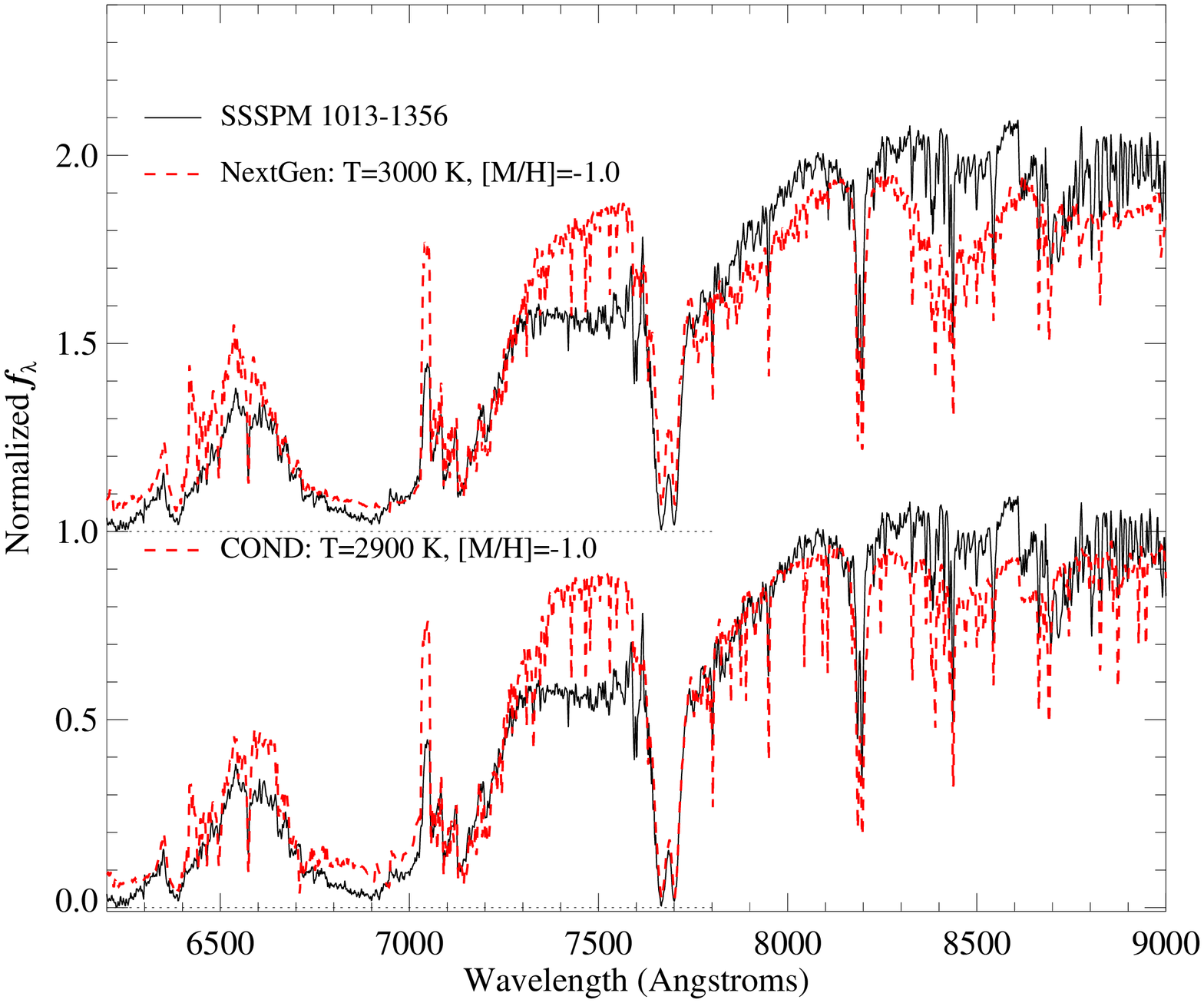}
\includegraphics[width=0.48\textwidth]{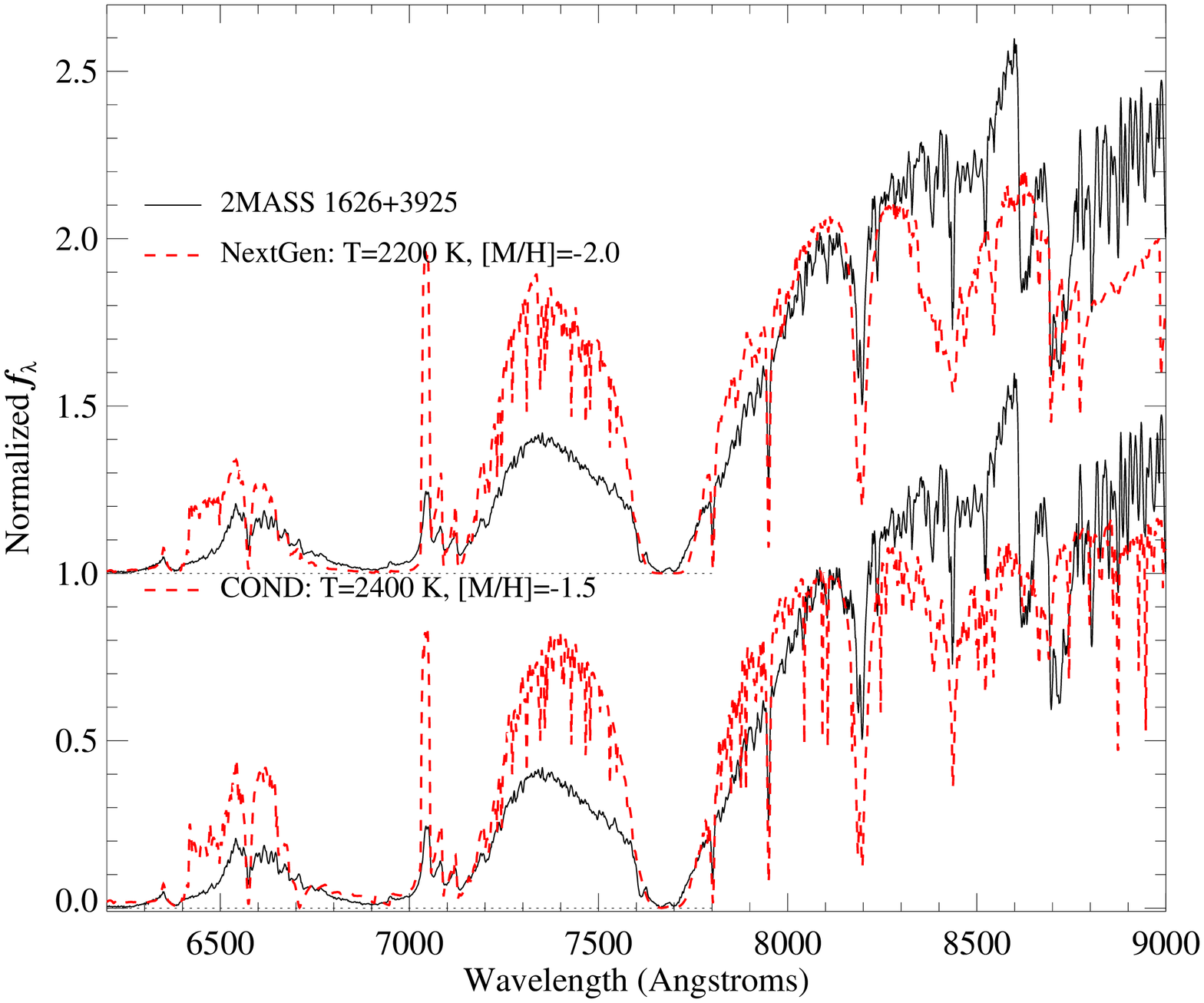}
\caption{Same as Figure~\ref{fig_modelfits} for the subdwarfs
2MASS 0142+0523, 2MASS 1013-1356 and 2MASS 1626+3925.
\label{fig_modelfits2}}
\end{figure}

\begin{figure}
\centering
\epsscale{0.8}
\plotone{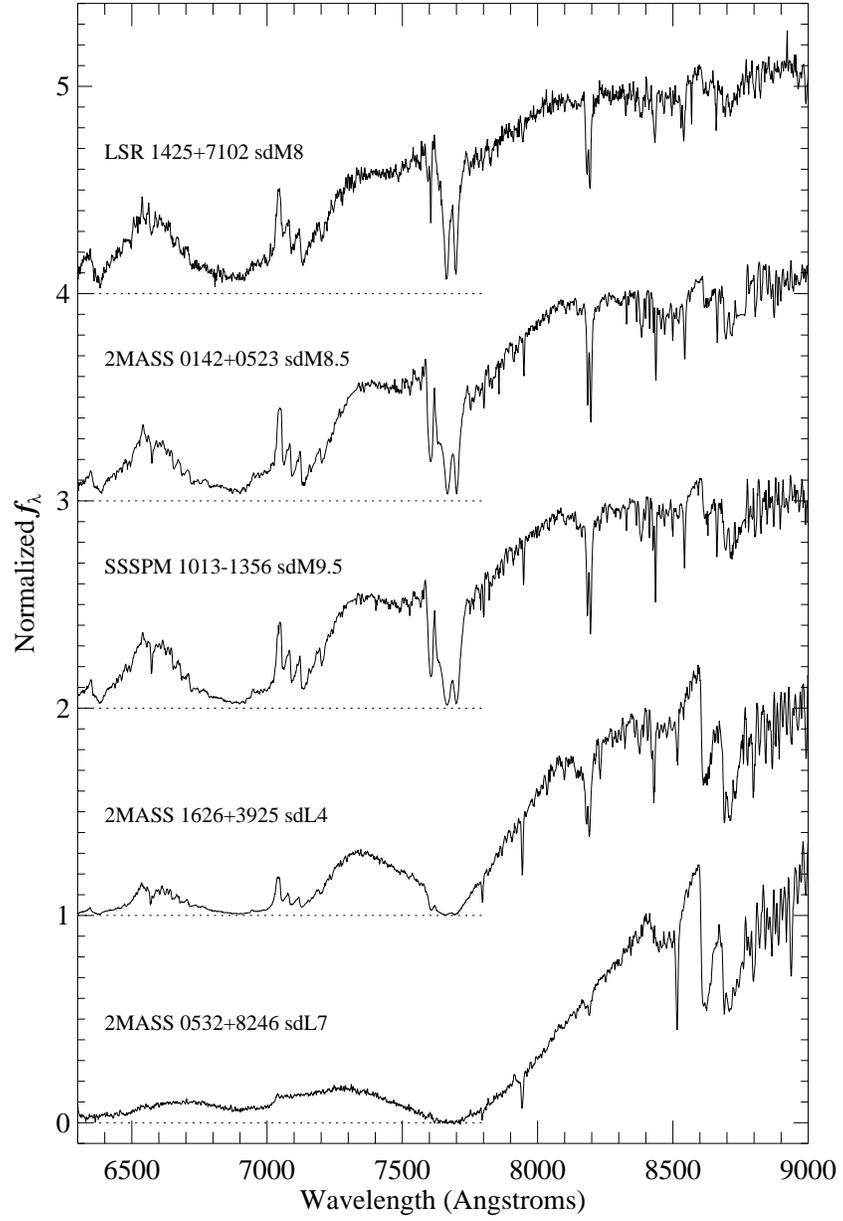}
\caption{Optical spectra sequence of five late-type ultracool subdwarfs, 
from top to bottom: LSR 1425+7102 (sdM8, data from \citet{lep1425}),
2MASS 0142+0523 (sdM8.5), SSSPM 1013-1356 (sdM9.5),
2MASS 1626+3925 (sdL4) and 2MASS 0532+8246 (sdL7; data from \citet{me0532}).
All data are shifted to their rest frame velocities,
normalized at 8350 {\AA}
and offset by a constant (dotted lines).   Data for 
LSR 1425+7102 have been corrected for telluric absorption. 
\label{fig_latesd}}
\end{figure}


\begin{thebibliography}{}


\bibitem[Abell(1959)]{abe59}Abell, G.\ O. 1959, \pasp, 67, 258

\bibitem[Ackerman \& Marley(2001)]{ack01}Ackerman, A.\ S., \& Marley, M.\ S.
2001, \apj, 556, 872

\bibitem[Allard et al.(2001)]{all01}Allard, F., Hauschildt, P.\ H.,
Alexander, D.\ R., Tamanai, A., \&  Schweitzer, A. 2001, \apj, 556, 357

\bibitem[Allington-Smith et al.(1994)]{all94} Allington-Smith, J., et al. 1994, 
\pasp, 106, 983

\bibitem[Artigau et al.(2006)]{art06}Artigau, E., Doyon, R., Lafreniere, D.,
Nadeau, D., Robert, J., \& Albert, L. 2006, \apj, in press

\bibitem[Bessell(1982)]{bes82}Bessell, M.\ S. 1982, PASA, 4, 417

\bibitem[Bessell \& Brett(1988)]{bes88}Bessell, M.\ S., \& Brett, J.\ M.
1988, \pasp, 100, 1134

\bibitem[Borysow, J{\o}rgensen, \& Zheng(1997)]{bor97}Borysow, A., J{\o}rgensen,
U.\ G., \& Zheng, C. 1997, \aap, 324, 185

\bibitem[Burgasser(2004)]{me1626}Burgasser, A.\ J. 2004 \apj, 614, L73

\bibitem[Burgasser \& Kirkpatrick(2006)]{melehpm2-59}Burgasser, A.\ J., \& Kirkpatrick, J.\ D. 2006, \apj, 645, 1485

\bibitem[Burgasser, Kirkpatrick \& Burrows(2006)]{metgrav}Burgasser, A.\ J., Kirkpatrick,
J.\ D., \& Burrows, A. 2006, \apj, 639, 1095

\bibitem[Burgasser et al.(2003a)]{me0532}Burgasser, A.\ J., Kirkpatrick,
J.\ D., Burrows, A., Liebert, J., Reid, I.\ N., Gizis, J.\ E., McGovern, M.\ R.,
Prato, L., \& McLean, I.\ S. 2003a, \apj, 592, 1186

\bibitem[Burgasser, Kirkpatrick, \& L\'epine(2005)]{mecs13}Burgasser, A.\ J., Kirkpatrick, J.\ D., \& L\'epine, S. 2005 in The 13th Cambridge Workshop on Cool Stars, Stellar Systems, and the Sun (ESA-SP-560), ed.\ F.\ Favata, G.\ A.\ J.\ Hussain \& B.\ Battrick (Noordwijk: ESA), p.\ 237

\bibitem[Burgasser et al.(2003)]{me03opt}Burgasser, A.\ J., Kirkpatrick,
J.\ D., Liebert, J., \& Burrows, A. 2003, \apj, 594, 510

\bibitem[Burgasser et al.(2003)]{mewide1}Burgasser, A.\ J., Kirkpatrick,
J.\ D., McElwain, M.\ W., Cutri, R.\ M., Burgasser, A.\ J., \& Skrutskie, M.\ F.
2003, \aj, 125, 850

\bibitem[Burgasser et al.(2004)]{mewide3}Burgasser, A.\ J., McElwain, M.\ W.,
Kirkpatrick, J.\ D., Cruz, K.\ L., Tinney, C.\ G., \& Reid, I.\ N. 2004 \aj,
127, 2856

\bibitem[Burgasser et al.(2002)]{me02a} Burgasser, A.\ J., et al. 2002, \apj, 564, 421

\bibitem[Burrows, Marley, \& Sharp(2000)]{bur00}Burrows, A., Marley, M.\ S.,
\& Sharp, C.\ M. 2000, \apj, 531, 438

\bibitem[Burrows \& Sharp(1999)]{bur99}Burrows, A., \& Sharp, C.\ M. 1999, \apj,
512, 843

\bibitem[Cannon(1984)]{can84}Cannon, R.\ D. 1984, in Astronomy with with Schmidt-Type Telescopes, Proc.\ IAU Coll.\ 78, ed.\ M.\ Cappaccioli 
(Dordrecht: Reidel), p.\ 25

\bibitem[Chiu et al.(2006)]{chi06}Chiu,, K., Fan, X., Leggett, S.\ K., Golimowski, D.\ A., 
Zheng, W., Geballe, T. R., Schneider, D. P., \& Brinkmann, J. 2006, \aj, 131, 2722 

\bibitem[Corbally, Gray \& Garrison(1994)]{cor94} Corbally, C.\ J., Gray R.\ O.,
\& Garrison, R.\ F. 1994, The MK process at 50 years. A Powerful Tool for Astrophysical Insight
(San Francisco: Astronomical Society of the Pacific)

\bibitem[Cruz et al.(2003)]{cru03} Cruz, K.\ L., Reid, I.\ N., Liebert, J., Kirkpatrick, J.\ D.,
\& Lowrance, P.\ J. 2003, AJ, 126, 2421

\bibitem[Cushing \& Vacca(2006)]{cus06}Cushing, M.\ C., \& Vacca, W.\ D. 2006, \aj, 131, 1797

\bibitem[Cutri et al.(2003)]{cut03}Cutri, R.\ M., et al. 2003,
\url{http://www.ipac.caltech.edu/2mass/releases/allsky/doc/explsup.html}

\bibitem[Dahn et al.(2002)]{dah02}Dahn, C.\ C., et al. 2002, \aj,
124, 1170

\bibitem[Dawson \& De Robertis(2000)]{daw00}Dawson, P.\ C., \& De Robertis,
M.\ M. 2000, \aj, 120, 1532

\bibitem[Deacon, Hambly \& Cooke(2005)]{dea05}Deacon, N.\ R., Hambly, N.\ C., \& 
Cooke, J.\ A. 2005, \aap, 435, 363

\bibitem[Dehnen \& Binney(1998)]{deh98}Dehnen, W., \& Binney, J.\ J. 1998, \mnras,
298, 387

\bibitem[Delfosse et al.(1997)]{del97}Delfosse, X., et al. 1997, \aap,
327, L25

\bibitem[Digby et al.(2003)]{dig03}Digby, A.\ P., Hambly, N.\ C., Cooke, J.\ A., Reid, I.\ N.,
\& Cannon, R.\ D. 2003, \mnras, 344, 583

\bibitem[Geballe et al.(2002)]{geb02}Geballe, T.\ R., et al. 2002, \apj, 564, 466

\bibitem[Gizis(1997)]{giz97}Gizis, J.\ E. 1997, \aj, 113, 806

\bibitem[Gizis \& Harvin(2007)]{giz07}Gizis, J.\ E., \& Harvin, J. 2007, \aj, in press

\bibitem[Gizis \& Reid(1997)]{giz97b}Gizis, J.\ E., \& Reid, I.\ N. 1997, \pasp, 109, 849

\bibitem[Hambly et al.(2001a)]{ham01a}Hambly, N.\ C., Davenhall, A.\ C.,
Irwin, M.\ J., \& MacGillivray, H.\ T. 2001a, MNRAS 326, 1315

\bibitem[Hambly et al.(2001b)]{ham01b}Hambly, N.\ C., Irwin, M.\ J., \& MacGillivray, H.\ T.
2001b, MNRAS 326, 1295

\bibitem[Hambly et al.(2001c)]{ham01c}Hambly, N.\ C., MacGillivray, H.\ T., Read, M.\ A.,
et al. 2001c, MNRAS 326, 1279

\bibitem[Hamuy et al.(1994)]{ham94}Hamuy, M., Suntzeff, N.\ B., Heathcote, S.\ R., Walker, A.\ R.,
Gigoux, P., \& Phillips, M.\ M. 1994, PASP, 106, 566

\bibitem[Hartley \& Dawe(1981)]{har81}Hartley, M., \& Dawe, J.\ A. 1981, PASA, 4, 251

\bibitem[Hauschildt, Allard \& Baron(1999)]{hau99}Hauschildt, P.\ H., Allard, F.,
\& Baron, E. 1999, \apj, 512, 377

\bibitem[Hawley, Gizis, \& Reid(1996)]{haw96}Hawley, S.\ L., Gizis, J.\ E.,
\& Reid, I.\ N. 1996, \aj, 112, 2799

\bibitem[Hawley et al.(1999)]{haw99}Hawley, S.\ L., Reid, I.\ N., Gizis, J.\ E.,
\& Byrne, P.\ B. 1999, in ASP Conf.\ Ser.\ 158, Solar and Stellar Activity: Similarities and Differences, ed.\ C.\ J.\ Butler \& J.\ G.\ Doyle (San Francisco: ASP), p.\ 63
\bibitem[Hawley et al.(2002)]{haw02}Hawley, S.\ L. et al. 2002, \aj, 123, 3409

\bibitem[Hook et al.(2004)]{hoo04}Hook, I., J{\o}rgensen, I., Allington-Smith, J.\ R., Davies, R.\ L., 
Metcalfe, N., Murowinski, R.\ G., \& Crampton, D. 2004, \pasp, 116, 425

\bibitem[Keenan \& McNeil(1976)]{kee76}Keenan, P.\ C., \& McNeil, R.\ C. 1976,
An Atlas of Spectra of the Cooler Stars: Types G, K, M, S and C (Columbus: Ohio State
Univ.\ Press)

\bibitem[Kirkpatrick, Henry, \& Irwin(1997)]{kir97}Kirkpatrick, J.\ D.,
Henry, T.\ J., \& Irwin, M.\ J. 1997, \aj, 113, 1421

\bibitem[Kirkpatrick, Henry, \& McCarthy(1991)]{kir91}Kirkpatrick, J.\ D.,
Henry, T.\ J., \& McCarthy, D.\ W., Jr. 1991, \apjs, 77, 417

\bibitem[Kirkpatrick et al.(2000)]{kir00}Kirkpatrick, J.\ D., Reid, I.\ N.,
Liebert, J., Gizis, J.\ E., Burgasser, A.\ J., Monet, D.\ G., Dahn, C.\ C.,
Nelson, B., \& Williams, R.\ J. 2000, \aj, 120, 447

\bibitem[Kirkpatrick et al.(1999)]{kir99}Kirkpatrick, J.\ D., et al. 1999,
\apj, 519, 802

\bibitem[Knapp et al.(2004)]{kna04}Knapp, G., et al. 2004, \apj, 127, 3553

\bibitem[Kurucz(1988)]{kur88}Kurucz, R.\ L. 1988, Trans.\ IAU, XXB, ed.\ M.\ McNally, (Dordrecht: Kluwer), p.\ 168-172

\bibitem[Kurucz \& Bell(1995)]{kur95}Kurucz, R.\ L., \& Bell, B. 1995, Atomic Line Data, Kurucz CD-ROM No. 23. (Cambridge: Smithsonian Astrophysical Observatory)

\bibitem[Leggett et al.(2000)]{leg00}Leggett, S.\ K., Allard, F., Dahn, C.,
Hauschildt, P.\ H., Kerr, T.\ H., \& Rayner, J. 2000, \apj, 535, 965

\bibitem[Leggett, Allard, \& Hauschildt(1998)]{leg98}Leggett, S.\ K.,
Allard, F., \& Hauschildt, P.\ H. 1998, \apj, 509, 836

\bibitem[L\'epine, Rich, \& Shara(2003a)]{lep03} L\'epine, S.,
Rich, R.\ M., \& Shara, M.\ M. 2003a, \aj, 125, 1598 

\bibitem[L\'epine, Rich, \& Shara(2003b)]{lep1610} L\'epine, S.,
Rich, R.\ M., \& Shara, M.\ M. 2003b, \apj, 591, L49

\bibitem[L\'epine et al.(2002)]{lep1826} L\'epine, S.,
Rich, R.\ M., Neill, J.\ D., Caulet, A., \& Shara, M.\ M. 2002, \apj, 581, L47

\bibitem[L\'epine \& Shara(2005)]{lep05} L\'epine, S., \& Shara, M.\ M. 2005, \aj, 129, 1483

\bibitem[L\'epine, Shara, \& Rich(2002)]{lep02} L\'epine, S., Shara, M.\ M., \&
Rich, R.\ M. 2002, \aj, 124, 1190

\bibitem[L\'epine, Shara, \& Rich(2003)]{lep1425} L\'epine, S., Shara, M.\ M., \&
Rich, R.\ M. 2003, \apj, 585, L69

\bibitem[L\'epine, Shara, \& Rich(2004)]{lep0822} L\'epine, S., Shara, M.\ M., \&
Rich, R.\ M. 2004, \apj, 602, L125

\bibitem[Liebert \& Probst(1987)]{lie87}Liebert, J., \& Probst, R.\ G. 1987, \araa, 25, 473

\bibitem[Linsky(1969)]{lin69}Linsky, J.\ L. 1969, \apj, 156, 989

\bibitem[Lodders(2002)]{lod02}Lodders, K. 2002, \apj, 577, 974

\bibitem[Lodieu et al.(2005)]{lod05}Lodieu, N., Scholz, R.-D., McCaughrean, M.\ J., Ibata, R., Irwin, M., \& Zinnecker, H. 2005, \aap, 440, 1061

\bibitem[Luyten(1979a)]{luy79a}Luyten, W.\ J. 1979a, LHS Catalogue: A Catalogue of Stars with Proper
Motions Exceeding 0$\farcs$5 Annually (Minneapolis: Univ.\ Minn.\ Press)

\bibitem[Martin, Fuhr \& Wiese(1988)]{mar88}Martin, G\ A., Fuhr, J.\ R., \& Wiese, W.\ L. 
1988, J.\ Phys.\ Chem.\ Ref.\ Data Suppl., 17, 3

\bibitem[Massey \& Gronwall(1990)]{mas90}Massey, P., \& Gronwall, C. 1990, \apj, 358, 344

\bibitem[Massey et al.(1988)]{mas88}Massey P., Strobel, K., Barnes, J.\ V., \& Anderson, E. 1988, \apj, 328, 315

\bibitem[Monet et al.(1992)]{mon92}Monet, D.\ G., Dahn, C.\ C., Vrba, F.\ J.,
Harris, H.\ C., Pier, J.\ R., Luginbuhl, C.\ B., \& Ables, H.\ D. 1992,
\aj, 103, 638

\bibitem[Monet et al.(1998)]{mon98}Monet, D.\ G., et al.\ 1998,
USNO-A2.0 Catalog (Flagstaff: USNO)

\bibitem[Morgan \& Keenan(1973)]{mor73} Morgan, W.\ W., \&
Keenan, P.\ C. 1973, \araa, 11, 29

\bibitem[Morgan, Keenan \& Kellman(1943)]{mor43} Morgan, W.\ W.,
Keenan, P.\ C., \& Kellman, E. 1943, An Atlas
of Stellar Spectra, with an Outline of Spectral Classification
(Chicago: Univ.\ Chicago Press)

\bibitem[Morgan et al.(1992)]{mor92}Morgan, D.\ H., Tritton, S.\ B., Savage, A., Hartley, M., \&
Cannon, R.\ D. 1992, in Digitised Optical Sky Surveys, ed. H.\ T.\ MacGillivray \&
E.\ B.\ Thomson (Dordrecht: Boston), p.\ 11

\bibitem[Mould \& Hyland(1976)]{mou76}Mould, J.\ R., \& Hyland, A.\ R. 1976, \apj, 208, 399

\bibitem[Mould \& McElroy(1976)]{mou78}Mould, J.\ R., \& McElroy, D.\ B. 1978, \apj, 220, 935

\bibitem[Pokorny et al.(2004)]{por04}Pokorny, R.\ S., Jones, H.\ R.\ A.,
Hambly, N.\ C., \& Pinfield, D.\ J. 2004, \aap, 421, 763

\bibitem[Rebolo, Mart{\'{\i}}n, \& Magazzu(1992)]{reb92}Rebolo, R.,
Mart{\'{\i}}n, E.\ L., \& Magazzu, A. 1992, \apj, 389, L83

\bibitem[Reid(2003)]{rei03}Reid, N. 2003, \mnras, 342, 837

\bibitem[Reid et al.(2000)]{rei00}Reid, I.\ N., Kirkpatrick, J.\ D.,
Gizis, J.\ E., Dahn, C.\ C., Monet, D.\ G., Williams, R.\ J.,
Liebert, J., \& Burgasser, A.\ J. 2000, \aj, 119, 369

\bibitem[Reid \& Gizis(2005)]{rei05}Reid, I.\ N., \& Gizis, J.\ E.
2005, \pasp, 117, 676

\bibitem[Reid, Hawley, \& Gizis(1995)]{rei95}Reid, I.\ N., Hawley, S.\ L.,
\& Gizis, J.\ E.  1995, \aj, 110, 1838

\bibitem[Reid et al.(2002)]{rei02}Reid, I.\ N., Kirkpatrick, J.\ D.,
Liebert, J., Gizis, J.\ E., Dahn, C.\ C., \&
Monet, D.\ G. 2002, \aj, 124, 519

\bibitem[Reid et al.(1991)]{rei91}Reid, I.\ N., et al. 1991, \pasp, 103, 661

\bibitem[Reiners \& Basri(2006)]{rei06} Reiners, A., \& Basri, G. 2006, \aj, 131, 1806

\bibitem[Saumon et al.(1994)]{sau94}Saumon, D., Bergeron, P., Lunine, J.\ I.,
Hubbard, W.\ B., \& Burrows, A. 1994, \apj, 424, 333

\bibitem[Scholz et al.(2000)]{sch00} Scholz, R.-D., Irwin, M., Ibata, R.,
Jahrei$\beta$, H., \& Malkov, O.\ Yu. 2000, \aap, 353, 958

\bibitem[Scholz et al.(2004a)]{sch1013} Scholz, R.-D., Lehmann, I., Matute, I.,
\& Zinnecker, H. 2004, \aap, 425, 519

\bibitem[Scholz et al.(2004b)]{sch1444} Scholz, R.-D., Lodieu, J., \& McCaughrean, M. 2004, \aap, 428, L25 

\bibitem[Sivarani, Kembhavi \& Gupchup(2006)]{siv06} Sivarani, T., Kembhavi, A.\ K., \& 
Gupchup, J. 2006, \apj, submitted

\bibitem[Schweitzer et al.(1999)]{sch99}Schweitzer, A., Shultz, R.-D., Stauffer, J.,
Irwin, M., \& McCaughren, M.\ J. 1999, \aap, 350, L62

\bibitem[Skrutskie et al.(2006)]{skr06}Skrutskie, M.\ F., et al. 2006, \aj, 131, 1163

\bibitem[Sommer-Larsen \& Zhen(1990)]{som90}Sommer-Larsen, J., \& Zhen, C. 1990, 
\mnras, 242, 10

\bibitem[Str\"{o}mgren(1987)]{str87}Str\"{o}mgren, B. 1987, in The Galaxy, ed.\ G.\ Gilmore 
\& B.\ Carswell (Dordrecht: Reidel), p.\ 299

\bibitem[Tsuji, Ohnaka, \& Aoki(1996a)]{tsu96a}Tsuji, T., Ohnaka, K., \&
Aoki, W. 1996, \aap, 305, L1

\bibitem[Tsuji et al.(1996)]{tsu96b}Tsuji, T., Ohnaka, K.,
Aoki, W., \& Nakajima, T. 1996, \aap, 308, L29

\bibitem[Vrba et al.(2004)]{vrb04}Vrba, F.\ J., et al. 2004, \aj, 127, 2948

\end{thebibliography}
\end{document}